%% file: 0_main.tex
\documentclass[sigconf]{acmart}

\usepackage{booktabs}
\usepackage{multirow}
\usepackage{pifont}
\usepackage{url}
\usepackage{xcolor}
\usepackage{hyperref}
\usepackage{graphicx}
\usepackage{subcaption}

\usepackage{listings}

\usepackage{makecell}
\usepackage{enumitem}

\AtBeginDocument{%
  }

\copyrightyear{2026}
\acmYear{2026}
\setcopyright{cc}
\setcctype{by}
\acmConference[ICSE '26]{2026 IEEE/ACM 48th International Conference on Software Engineering}{April 12--18, 2026}{Rio de Janeiro, Brazil}
\acmBooktitle{2026 IEEE/ACM 48th International Conference on Software Engineering (ICSE '26), April 12--18, 2026, Rio de Janeiro, Brazil}
\acmPrice{}
\acmDOI{10.1145/3744916.3787792}
\acmISBN{979-8-4007-2025-3/2026/04}

\newcommand{\toolname}{\textsc{CoHiKer}}

\definecolor{oldgray}{gray}{0.55}
\definecolor{newblue}{rgb}{0.1, 0.3, 0.9}
\newcommand{\old}[1]{}

\definecolor{reviewpurple}{HTML}{6A0DAD}

\begin{document}

\title{
Towards Better Linux Kernel Fault Localization: Leveraging Contrastive Reasoning and Hierarchical Context Analysis
}

\author{Haichi Wang}
\email{wanghaichi@tju.edu.cn}
\orcid{0009-0007-6953-8369}
\affiliation{%
  \institution{School of Computer Software, Tianjin University}
  \city{Tianjin}
  \country{China}
}

\author{Ruiguo Yu}
\email{rgyu@tju.edu.cn}
\orcid{0000-0003-4039-2355}
\affiliation{%
  \institution{School of Computer Science and Technology, Information and Network Center, Tianjin University}
  \city{Tianjin}
  \country{China}
}

\author{Yesong Pang}
\email{pangyesong_0503@tju.edu.cn}
\orcid{0009-0008-3132-045X}
\affiliation{%
  \institution{School of Computer Software, Tianjin University}
  \city{Tianjin}
  \country{China}
}

\author{Yingquan Zhao}
\email{zhaoyingquan@tju.edu.cn}
\orcid{0000-0003-2998-1052}
\affiliation{%
  \institution{School of Cybersecurity, Tianjin University}
  \city{Tianjin}
  \country{China}
}

\author{Junjie Chen}
\authornote{Junjie Chen is the corresponding author.}
\email{junjiechen@tju.edu.cn}
\orcid{0000-0003-3056-9962}
\affiliation{%
  \institution{School of Computer Software, Tianjin University}
  \city{Tianjin}
  \country{China}
}

\author{Jiajun Jiang}
\email{jiangjiajun@tju.edu.cn}
\orcid{0000-0003-1983-6572}
\affiliation{%
  \institution{School of Computer Software, Tianjin University}
  \city{Tianjin}
  \country{China}
}

\author{Zan Wang}
\email{wangzan@tju.edu.cn}
\orcid{0000-0001-6173-8170}
\affiliation{%
  \institution{School of Computer Software, School of Artificial Intelligence, Tianjin University}
  \city{Tianjin}
  \country{China}
}

\begin{abstract}

Debugging the Linux kernel remains a formidable challenge due to its vast codebase, complex architecture, and low-level programming intricacies. Effective fault localization (FL) is thus essential for efficient kernel debugging and maintenance. While existing FL techniques (both traditional and LLM-based) have shown promise in general-purpose software, they are ill-suited for the kernel context. In particular, recent LLM-based techniques often treat bug reports and source code as plain text, lacking deep integration of kernel-specific knowledge, which limits their ability to identify root causes and achieve fine-grained localization.

We present \toolname{}, a novel LLM-based FL technique tailored to the Linux kernel. \toolname{} introduces two key innovations: (1) contrastive reasoning, which identifies root causes by analyzing the behavioral divergence between carefully mutated passing and failing test cases, and (2) hierarchical context analysis, which systematically narrows the localization scope from files to methods by integrating crash reports, syscall semantics, inter-file dependencies, and kernel-specific features. 
Unlike prior techniques that rely on static understanding and full-code input, \toolname{} decomposes the localization task and enables structured LLM prompting to reason semantically over meaningful contexts.

We evaluate \toolname{} on an extended Linux kernel bug dataset against five state-of-the-art baselines. \toolname{} consistently outperforms all competitors, improving $\textit{Top}$-$1$ localization accuracy by up to 26.07\% at the file level and 56.85\% at the method level over state-of-the-art LLM-based baselines, while achieving up to 8.84× and 28.9× reductions in token consumption, respectively\old{, highlighting its scalability and efficiency. These results demonstrate the promise of \toolname{} as a practical and accurate solution for kernel FL.}. Furthermore, \toolname{} demonstrates strong generalizability on the non-kernel dataset, with comparable gains (15.5\% and 5.3\% in $\textit{Top}$-$1$ at file and method levels). These results demonstrate the promise of \toolname{} as a practical and accurate solution for kernel FL.

\end{abstract}

\ccsdesc[500]{Software and its engineering~Software testing and debugging}

\keywords{Fault Localization, Linux Kernel, Large Language Model}

\maketitle

\setlength{\abovecaptionskip}{1.5mm}
\setlength{\belowcaptionskip}{-1.1mm}

\input{1_introduction}
\input{2_motivating_example}

\input{3_approach}
\input{5_evaluation}
\input{7_discussion}

\input{6_relatedwork}
\input{8_conclusion}

\input{10_acknowledge}

\newpage


\bibliographystyle{ACM-Reference-Format}
\bibliography{sample-base}











\end{document}

%% file: 1_introduction.tex
\section{Introduction}
\label{sec:introduction}

\newcommand{\flfile}{\textsc{FL@F}}
\newcommand{\flmethod}{\textsc{FL@M}}

The Linux kernel, a core component of Linux-based operating systems, manages critical resources across diverse platforms. Over decades of continuous development, it has evolved into one of the largest software systems, exceeding 40 million lines of code~\cite{linux_kernel_40m}. 
Maintaining this massive codebase is challenging, and despite extensive community contributions, bugs persist, often impacting system stability and security.
While researches have focused on detecting kernel bugs, particularly through advanced fuzzing techniques~\cite{sun2021healer, zhao2022statefuzz, yang2025kernelgpt}, developers still devote substantial effort to localizing root causes before fixing~\cite{wang2025empirical}.
However, Kernel fault localization is inherently more complex than in general software due to: 
(1) The pronounced gap between bug symptoms and root causes, driven by asynchronous execution and cross-subsystem coupling that delay and obscure bug manifestations.
(2) The massive and highly interdependent codebase, comprising numerous functionally diverse subsystems and far exceeding other software in scale (e.g., the already sophisticated GCC compiler has only about 15 million lines of code~\cite{DBLP:conf/issta/Even-MendozaSDC23}).
(3) The frequent loss of diagnostic information, resulting from system-wide failures inherent to kernel bugs that erase runtime states and hinder the collection of crucial runtime data (e.g., code coverage).



Fault localization (FL) techniques identify code locations responsible for software failures, aiding developers in diagnosing and resolving bugs more efficiently~\cite{wong2016survey}. 
In general software, FL techniques fall into three categories: spectrum-based (SBFL)~\cite{zou2019empirical, wen2019historical}, mutation-based (MBFL)~\cite{jia2010analysis, andrews2006using}, and information retrieval-based (IRFL) ~\cite{xia2023information, zhang2019finelocator, chen2021pathidea}. 
However, the unique characteristics of the Linux kernel significantly hinder the effectiveness of these traditional techniques. 
SBFL depends on runtime coverage, often unavailable in kernel crashes \cite{hasanov2024little}; 
MBFL becomes computationally infeasible due to costly recompilation per mutation; 
IRFL, which rely on lexical or semantic similarity between bug reports and source code, struggles to cope with the kernel’s low-level semantics and indirect relationship between bug descriptions and faulty code \cite{wang2025empirical}.


Large Language Models (LLMs) have recently shown promise for FL in general software~\cite{yang2024swe, qin2024agentfl, zhang2024autocoderover}. 
LLM-based techniques~\cite{xia2024agentless, zhou2025benchmarking, wu2023large, yang2024large} leverage LLMs to interpret  bug reports and reason over code semantics to identify likely fault locations. 
However, their effectiveness remains limited, as existing techniques (e.g., Agentless~\cite{xia2024agentless}) rely on static reasoning that naively combine heterogeneous inputs (e.g., bug reports, crash traces, and source code) into uniform representations.
Such static and simplified reasoning is inadequate for kernel fault localization because it: 
(1) Neglects the distinctive nature of kernel bugs, where the gap between bug symptoms and root causes (e.g., a file-system bug triggers a crash in memory management through shared states, with the faulty file absent from the bug report) makes reasoning solely over static information insufficient.
(2) Overlooks the massive scale and complex dependencies in kernel, where even partial code contexts can exceed the input capacity of LLMs.
To improve FL for the Linux kernel, LinuxFL+\cite{zhou2025benchmarking} was recently proposed as the first LLM-based technique tailored to this domain. 
However, LinuxFL+ incorporates only limited kernel-specific knowledge
(such as incorporating filenames extracted from bug-related emails to expand suspicious file sets) and does not explicitly address the limitations of static reasoning and scalability issues that persist in prior LLM-based techniques.
Therefore, a more effective and kernel-aware FL technique is still needed.

To address these limitations, we propose \textbf{\toolname{}} (\textbf{Co}ntrastive Reasoning and \textbf{Hi}erarchical Context Analysis for \textbf{Ker}nel Fault Localization), a novel LLM-based FL technique tailored to the Linux kernel. \toolname{} introduces two key innovations:
(1) \textbf{Contrastive Reasoning}: 
To address the limitation of static reasoning, \toolname{} performs contrastive reasoning to bridge the gap between bug symptoms and root causes in kernel fault localization. In particular, \toolname{} mutates failing tests to generate passing variants and leverages LLMs to reason over their execution results differences (i.e., pass or fail).
By identifying causal differences between test behaviors, it isolates 
fault-relevant kernel features and narrows the root cause search space.
This alleviates the coverage loss caused by kernel crashes, and mitigates the reliance of SBFL\cite{zou2019empirical, wen2019historical} and MBFL\cite{jia2010analysis, andrews2006using} on code coverage or costly recompilation after kernel code mutations.
(2) \textbf{Hierarchical Context Analysis}: 
To cope with the kernel's large codebase under LLM input constraints,
\toolname{} employs a hierarchical analysis framework that integrates context at both file and method granularity, enabling scalable reasoning and progressive refinement for more precise localization.
At the file level, it identifies Top-K suspicious files by analyzing textual cues from bug reports and inter-file dependencies. 
At the method level, it analyzes internal code structures, such as call hierarchies and parameter flows, to rank suspicious methods within those files.
Finally, contrastive reasoning results (i.e., explanations of how the bug is triggered)
are integrated across both levels to refine the fault ranking by aligning kernel semantics with root cause hypotheses.
Together, these components enable \toolname{} to overcome the limitations of prior techniques, offering a kernel-specific, causally grounded, and context-aware framework for more accurate and interpretable FL in complex kernel environments.

To evaluate \toolname{}, we extended an existing kernel FL dataset \cite{mathai2024kgym} to 210 verified instances, comparing it against two categories of baselines: 
(1) traditional FL techniques, including the file-level IR-based technique PathIdea~\cite{chen2021pathidea}, the method-level IR-based technique FineLocator~\cite{zhang2019finelocator} and the method-level trace-based technique CrashLocator~\cite{wu2014crashlocator}; 
and (2) state-of-the-art LLM-based techniques, namely Agentless~\cite{xia2024agentless}, LinuxFL+~\cite{zhou2025benchmarking} (extended from SWE-Agent~\cite{yang2024swe}) and SoapFL~\cite{qin2025s} (for non-kernel context only).
As only LinuxFL+ was designed for the Linux kernel, we adapted the other techniques to ensure compatibility with the kernel context.
Results show \toolname{} achieves significantly higher FL accuracy. 
For example, the traditional PathIdea achieves only 6.7\% Top-1 accuracy at the file level, while \toolname{} achieves 55.24\%.
The traditional FineLocator cannot identify any faulty method as Top-1, while \toolname{} achieves 32.86\% Top-1 accuracy at the method level.
Compared to LLM-based baselines (Agentless and LinuxFL+), \toolname{} improves Top-1 accuracy by 4.5\%--26.70\% at the file level and 32.71\%--56.85\% at the method level.

\toolname{} also demonstrates superior efficiency, reducing token consumption by 4.46$\times$--8.84$\times$ at file level and 21.92$\times$--28.90$\times$ at method level compared to baselines.
Furthermore, \toolname{} shows strong generalizability on non-kernel systems (SWE-bench-lite \cite{jimenez2024swebench}), achieving improvements over state-of-the-art baselines (e.g., 15.5\% and 5.3\% in Top@1 at file and method levels) that align with its gains in the kernel domains.
These results confirm \toolname{}'s consistent superiority across diverse systems, highlighting the broad applicability of its core methodologies.


\smallskip
To sum up, this work makes the following major contributions:
\begin{itemize}
    \item We propose \toolname{}, a novel LLM-based FL technique tailored to the Linux kernel, combining contrastive reasoning and hierarchical context analysis to address the unique challenges in complex kernel scenarios.
    
    \item We introduce a contrastive reasoning component that leverages LLMs to compare failing and passing test cases, enabling the identification of root causes that are tightly linked to kernel-level bug manifestations.

    \item We design a hierarchical context analysis framework, which discriminatively incorporates file-level and method-level bug signals while optimizing for LLM input efficiency and contextual relevance.

    \item We conduct extensive experiments on an extended kernel FL dataset, showing that \toolname{} achieves superior localization accuracy and significantly reduces token consumption compared to state-of-the-art LLM-based baselines.

\end{itemize}

%% file: 2_motivating_example.tex
\section{An Illustrative Example}
\label{sec:motivating}

\begin{figure}[t]
  \centering
  \begin{subfigure}{\linewidth}
    \centering
    \includegraphics[width=1.0\linewidth]{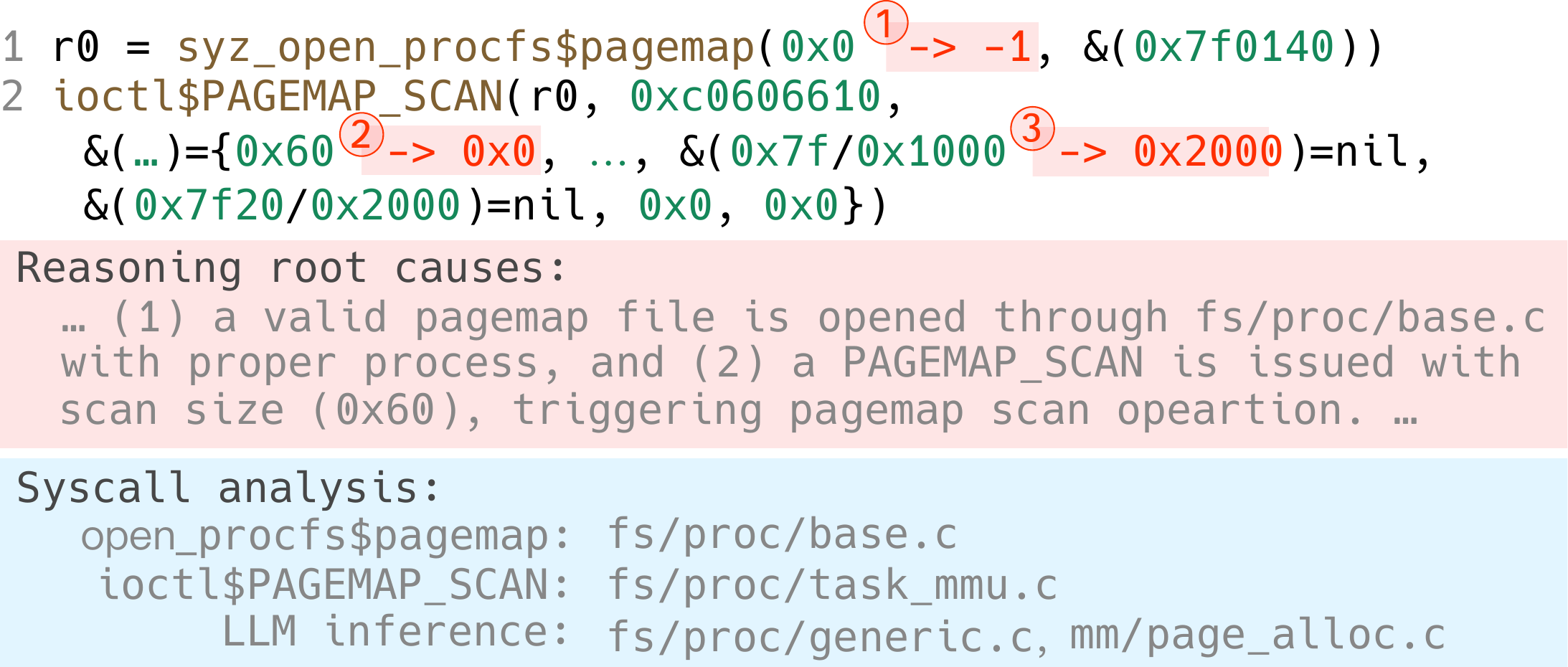}
    \caption{Test Case}
    \label{subfig:motivate-test}
  \end{subfigure}
  
  
  \begin{subfigure}{\linewidth}
    \centering
    \includegraphics[width=1.0\linewidth]{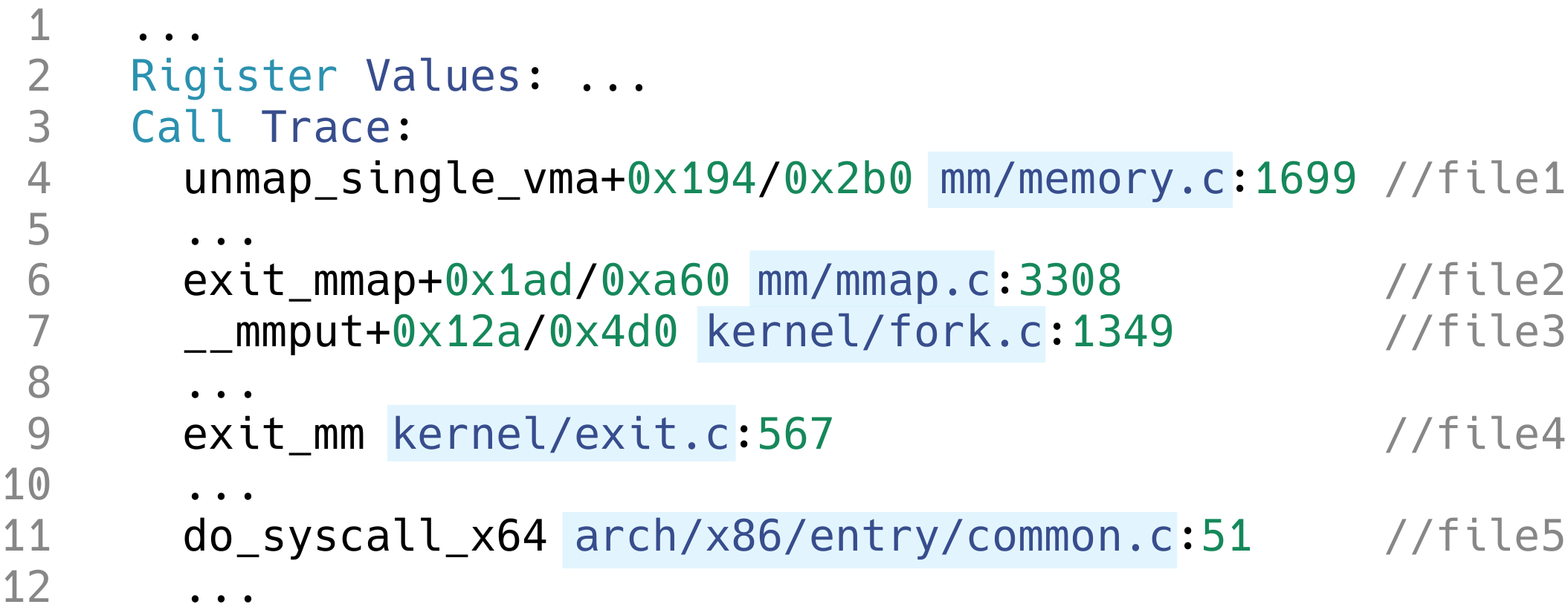}
    \caption{Crash Trace}
    \label{subfig:motivate-trace}
  \end{subfigure}
  
  \caption{Bug Report: WARNING in unmap\_page\_range \cite{syzbotcrash}}
  \label{fig:motivate}
\end{figure}

\begin{figure}[]
    \centering
    \includegraphics[width=1.0\linewidth]{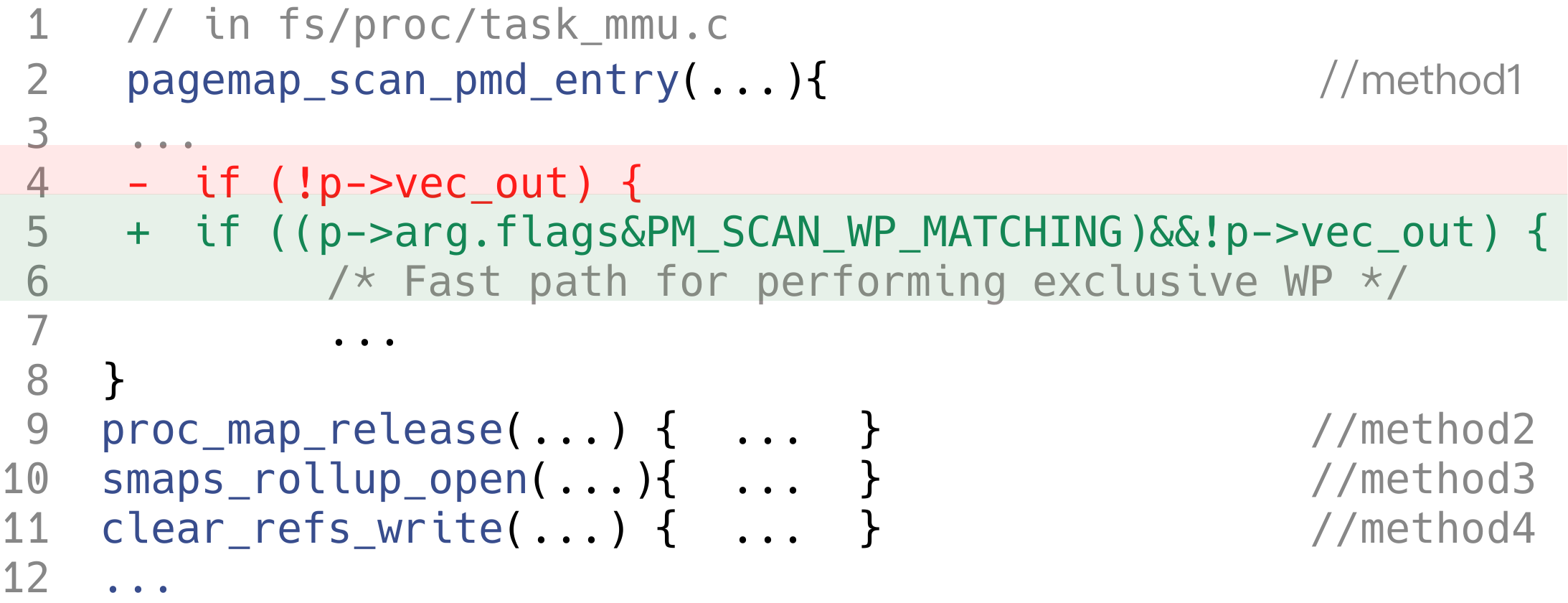}
    \caption{Suspicious File: fs/proc/task\_mmu.c}
        \label{fig:candidate-method}
\end{figure}

We use a real-world example \cite{syzbotcrash} to show \toolname{}'s kernel bug localization process.
Figure~\ref{subfig:motivate-test} shows a simplified test case that triggers a crash: it opens a file via \texttt{syz\_open\_procfs\$pagemap} (Line 1), then performs a \texttt{PAGEMAP\_SCAN} operation (Line 2). 
A bug in the kernel's memory management subsystem causes this operation to crash the system (\textit{ref.} Figure~\ref{fig:candidate-method}). 
Figure~\ref{subfig:motivate-trace} shows the corresponding crash trace (omitting repetitive frames).
While the trace captures the symptom, it does not clearly indicate the root cause, complicating localization given the semantic gap between the syscall-level test case and the underlying kernel code.

Specifically, the kernel crash precludes code coverage collection, making coverage-based techniques (e.g., SBFL) inapplicable.
Furthermore, the crash trace shows symptoms in the \texttt{mm/} and \texttt{kernel/} subsystems, but the root cause actually lies in the functionally distinct \texttt{proc/} subsystem. This semantic discrepancy between the crash context and the actual faulty location hinders IR-based methods from identifying correct files. While recent LLM-based methods attempt to bridge this gap by incorporating semantically relevant code, they often struggle with the significant functional misalignment between bug reports and the faulty code. Consequently, they either fail to include the desired faulty code or encounter a massive search space. For instance, LinuxFL+ still includes 3371 files in its refined space for this example, making it challenging to precisely localize the fault due to the limited LLM input capacity. Hence, more effective kernel fault localization techniques are urgently needed.

To address these limitations of existing approaches, \toolname{} begins with contrastive reasoning to uncover the root cause of the failure. 
This is done by comparing how small changes to the test case affect the observed execution results. 
Specifically, \toolname{} leverages the LLM (i.e., \textsc{DeepSeek-V3}\cite{deepseek_v3}) to generate mutated variants (three in this example) of the failing test case (Figure~\ref{fig:motivate}(a)) by slightly
altering each parameter (highlighted in red).
The LLM is guided to modify only interface-related arguments (e.g., flags, sizes) while preserving the syscall sequence, keeping each mutant semantically close to the original test.
Instead of treating each test in isolation, the LLM compares execution results across passing and failing tests and reasons about their implications.
Specifically, the first two mutations (\ding{172} and \ding{173}) prevent the test failure while the third (\ding{174}) does not. Using this contrast, along with an analysis of the passing/failing test cases and the syscall definition (explaining the functionality of each parameter), the LLM infers the root cause: faulty handling of valid input combinations in the kernel's memory management logic, specifically for a \texttt{PAGEMAP\_SCAN} operation using a file like 0x0 and a scan size of 0x60.
The LLM's reasoning output in Figure~\ref{subfig:motivate-test} demonstrates that by focusing on how test cases reveal or suppress faulty kernel behavior, \toolname{} enables deep, LLM-driven causal reasoning that 
isolates the true root causes of bugs. 

However, root cause inference alone is insufficient for actionable localization given the massive codebase in the kernel, such as still 3371 files left even by using the state-of-the-art LinuxFL+ for space refinement as aforementioned, making the fine-grained method-level fault localization extremely hard.
To make the process scalable and precise, \toolname{} applies hierarchical localization: first identifying a small set of suspicious files, then drilling down to individual methods.
At the file level, \toolname{} parses the crash trace (Figure~\ref{subfig:motivate-trace}) to extract five candidate files (marked in blue)
and analyzes system calls (e.g., {\tt ioctl\$PAGEMAP\_SCAN}) in the test case to infer more relevant files (e.g., {\tt fs/proc/task\_mmu.c}). The objective for this process is to create a manageable search space that includes the faulty files. It starts with the stack trace because the actual faulty files, even though not explicitly listed, must have dependencies with some file presented in the trace.
Specifically, \toolname{} leverages the LLM to infer additional dependent files based on test case semantics and crash context.
Finally, it ranks candidate files by aligning them with the contrastive reasoning output, which substantially narrows the search space.
In this case, as the bug involves faulty handling of \texttt{PAGEMAP\_SCAN}, \toolname{} forms a focused set of 18 files closely related to the \texttt{proc/} subsystem and
correctly prioritizes {\tt fs/proc/task\_mmu.c}, as shown in Figure~\ref{fig:candidate-method}.

\toolname{} then conducts method-level analysis within the set of suspicious files, as file-level granularity is often too coarse for practical debugging. 
The file {\tt fs/proc/task\_mmu.c} alone contains 96 methods (many omitted in Figure~\ref{fig:candidate-method} for brevity). 
As a consequence, it still results in 1,034 methods to analyze in this case even if only considering the top 10 suspicious files from the file-level results like existing approaches~\cite{zhou2025benchmarking}, which remains a substantial search space.
To further narrow the search space, \toolname{} prunes candidates with low likelihood of being faulty, using a defect prediction model trained on historical kernel bugs, which results in 81 methods left as suspicious.
For each remaining method, it prompts the LLM to summarize its functionality based on the method's signature and body.
It then ranks methods by matching summaries with the inferred root cause.
In this example, {\tt pagemap\_scan\_pmd\_entry} is ranked top-1, as its functionality closely matches the inferred root cause. 
This diagnosis is confirmed by the official patch, which modifies this method directly (Lines 4–6, Figure~\ref{fig:candidate-method}).

To sum up, \toolname{} enables a more accurate root cause inference of the fault and a more precise search space refinement at both file and method levels by integrating the contrastive reasoning and hierarchical context analysis,
addressing key 
limitations of prior techniques in the complex landscape of Linux kernel debugging.

%% file: 3_approach.tex
\section{Technique}
\label{sec:approach}

\begin{figure*}[htbp]
    \centering
    \includegraphics[width=0.95\linewidth]{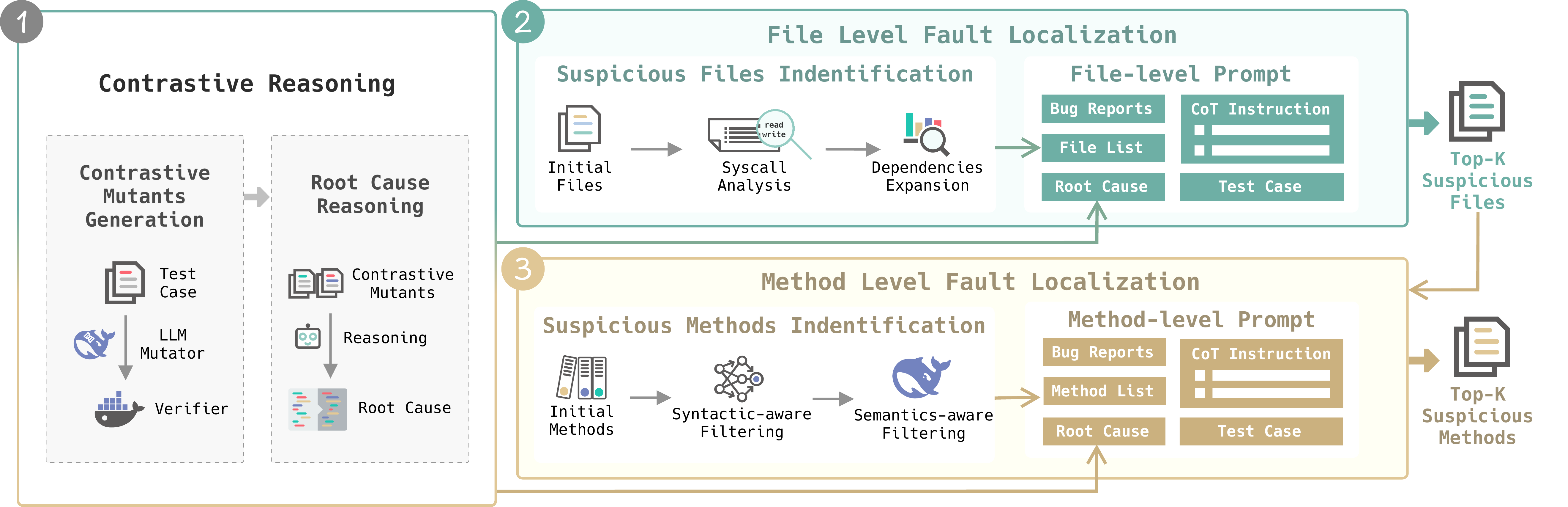}
    \caption{Overview of \toolname{}}
        \label{fig:overview}
\end{figure*}

In this paper, we propose \textbf{\toolname{}}, a framework for Linux kernel fault localization that combines \textit{contrastive reasoning} for root cause identification with \textit{hierarchical context analysis} for efficient diagnostic signal utilization within LLM input constraints.


Figure~\ref{fig:overview} illustrates \toolname{}'s three main stages: 
\textbf{First}, it leverages LLMs to identify precise root causes by analyzing both failing and intentionally generated passing test cases.
\textbf{Second}, it locates suspicious files via file-level fault localization (FL@F, stage \ding{173}).
Specifically, \toolname{} first identifies suspicious files using bug reports and system call (i.e., syscall) analysis via LLM-based context enrichment. 
The final candidates undergo LLM analysis to rank the Top-K most suspicious files by aligning contrastive reasoning results.
\textbf{Third}, it locates suspicious methods through method-level fault localization(FL@M, stage \ding{174}).
Specifically, given suspicious files by \flfile{}, \toolname{} combines a fault prediction model and LLMs to first identify the most probable methods, which are then analyzed by the LLM to rank and select the Top-K most suspicious methods via aligning the contrastive reasoning results.

\subsection{Contrastive Reasoning}
\label{subsec:cr}

As outlined in Section~\ref{sec:introduction}, precise root-cause analysis is essential for effective fault localization.
To this end, \toolname{} introduces a contrastive reasoning component that identifies bug-related root causes by analyzing test cases along with their execution outcomes, operating in two stages: contrastive mutants generation and root cause reasoning.

\begin{figure}[t]
    \centering
    \includegraphics[width=1.0\linewidth]{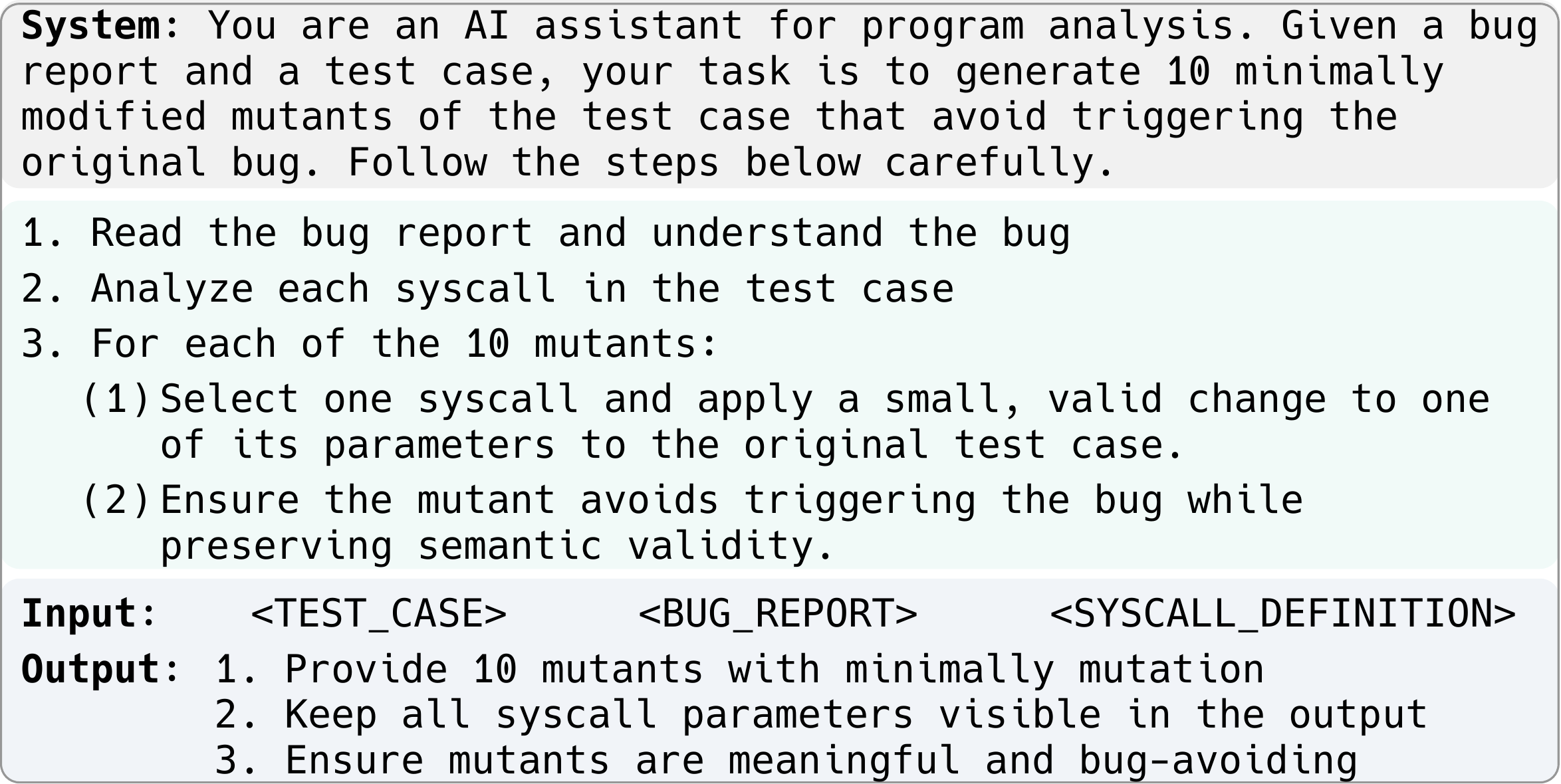}
    \caption{Prompt for Test Case Mutation}
        \label{fig:prompt-mutation}
\end{figure}

\toolname{} begins by leveraging mutation analysis~\cite{jia2010analysis} to generate contrastive mutants. 
Traditionally used to assess test suite quality, mutation analysis has also been applied to detect bugs in compilers and other systems\cite{wang2020deep, zhao2022history, dakhel2024effective}.
In \toolname{}, however, mutation analysis serves a different goal: to identify bug-relevant elements in test cases by observing which mutations cause the failure to disappear (i.e., pass) or persist (i.e., fail).
Given complex syscall parameter constraints, traditional handcrafted mutation operators struggle to preserve semantic correctness or guide mutations toward passing behavior.
Hence, \toolname{}'s LLM-based mutator leverages semantic understanding and contextual reasoning to generate minimally constraint-aware test variants.
The prompt (Figure~\ref{fig:prompt-mutation}) used in \toolname{} consists of three components: a role-defining system message, a specific task description, and an input-output format.
Following the common LLM prompting practice~\cite{yang2024swe}, \toolname{} adopts a Chain-of-Thought (CoT) prompting strategy to encourage step-by-step reasoning.

For a failing test case, the LLM is prompted to (1) analyze how the bug is triggered, (2) identify potential fault-inducing system call and its parameters, and (3) suggest minimal modifications to these parameters (e.g., file descriptors, pointers)
that may prevent the bug from being triggered.
\toolname{} then generates mutants with small syscall parameter changes.
This design choice is motivated by two reasons:
(1) Coarse-grained mutations (e.g., modifying entire syscall instructions) often break data dependencies (e.g., variable references), making it harder to isolate fault-relevant behavior.
(2) Fine-grained parameter-level mutations yield more precise causal signals, as parameters typically map to specific functionalities in kernel code and require fewer semantic checks.
Each generated mutant is executed to determine whether the original bug persists.
This process continues until a sufficient contrastive set is obtained (three passing mutants along with the original failing test by default, balancing LLM input constraints with reasoning quality in follow-up root cause reasoning).

\begin{figure}[t]
    \centering
    \includegraphics[width=1.0\linewidth]{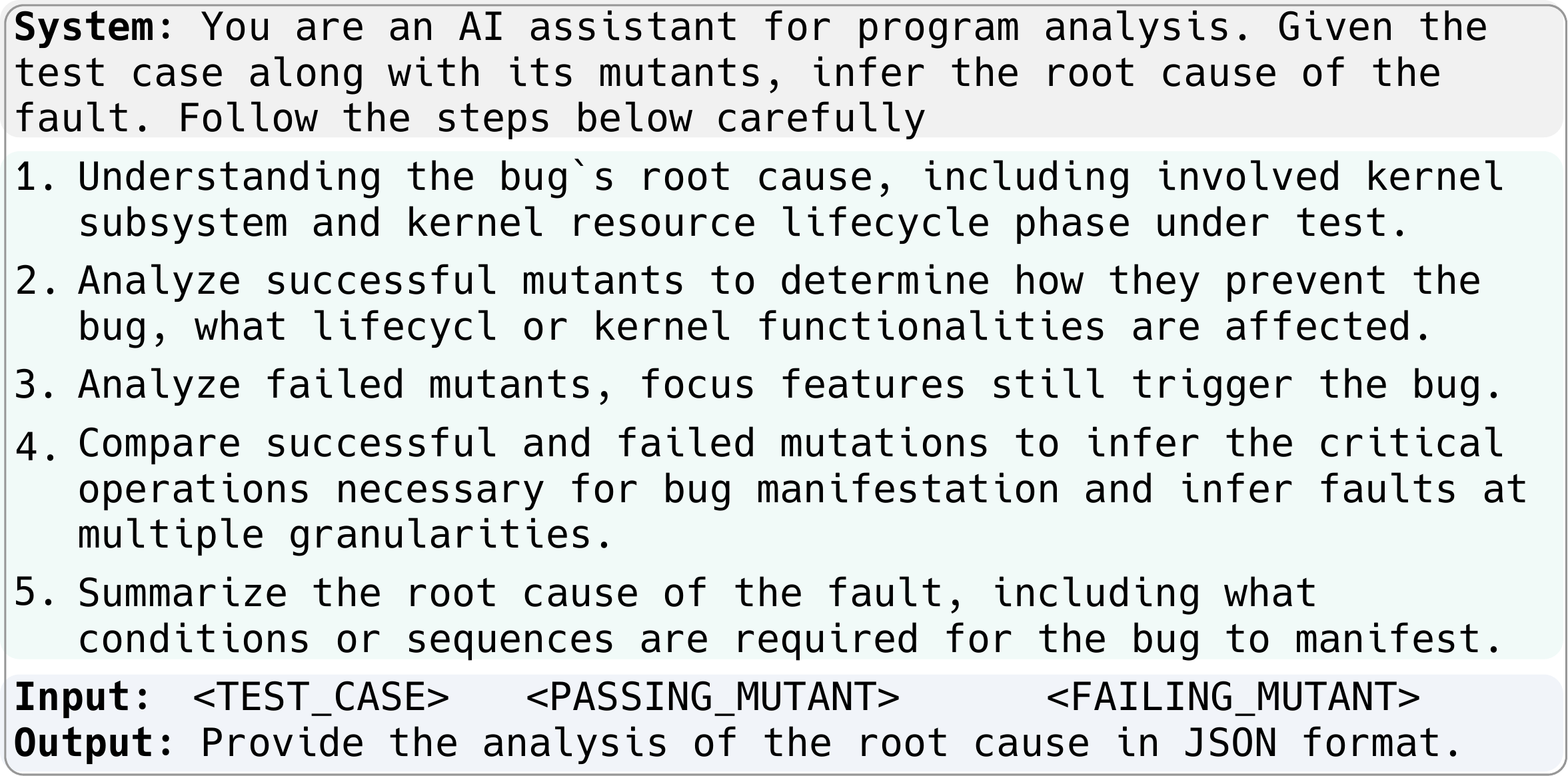}
    \caption{Prompt for Contrastive Reasoning}
        \label{fig:prompt-reasoning}
\end{figure}

With these contrastive mutants, \toolname{} infers the bug's root cause by comparing semantic differences between the original test case and its contrastive variants. 
This step also leverages LLMs, using a dedicated reasoning prompt shown in Figure~\ref{fig:prompt-reasoning}. 
The reasoning process unfolds in three stages:
(1) The LLM interprets the original test to infer exercised kernel subsystems (e.g., memory management, file systems).
(2) It then examines passing mutants to understand how specific input modifications may prevent the bug, thereby identifying the functionalities or subsystems affected by those changes.
(3) Finally, the LLM compares this analysis with the failing test to isolate the key differences that consistently correlate with bug manifestation.
By triangulating these insights, \toolname{} infers a precise and interpretable root cause that reflects the kernel-level logic responsible for the failure. 
This contrastive reasoning component allows \toolname{} to move beyond surface-level pattern matching and instead derive causal explanations grounded in system behavior.

\subsection{File-Level Fault Localization}
\label{subsec:flfl}

While contrastive reasoning infers plausible root causes, locating these within the kernel's complex codebase still remains challenging.
To address this, \toolname{} introduces a hierarchical context analysis framework that performs fault localization at both the file and method levels.
By leveraging the distinct semantics and granularity of each level, this design improves information utilization while narrowing the search space for fault localization.
Specifically, \toolname{} begins by identifying suspicious files, which then form the basis for finer-grained localization at the method level.

At the file level, \toolname{} begins by analyzing the crash report, which typically reveals the dynamic execution path taken during the failure. 
Kernel files referenced in the crash trace are treated as initial candidates, as they are likely involved in the observed malfunction. 
However, crash traces often provide only partial clues: asynchronous faults, delayed state corruptions, or silent failures may prevent key faulty files from appearing in the trace.
To address this limitation, \toolname{} complements the crash report with a deeper analysis of the failing test case to identify additional fault-relevant files that may not surface in the trace. 
Specifically, it expands candidates from two perspectives:
(1) \textbf{System Call Analysis}: 
\toolname{} leverages LLMs to identify syscalls in the test case and map them to 
semantically related kernel files, where static analysis is often unreliable and fails under heavy macro expansion and conditional compilation \cite{tartler2014static}.
For instance, a file-open syscall may 
span file-system and memory-management in ways obscured by macro expansion, causing static analysis to miss these links, whereas LLM can bridge the gap by leveraging semantic cues from naming, documentation, and usage contexts.
(2) \textbf{Dependency Expansion}: \toolname{} further uses LLMs to infer semantic and functional dependencies among kernel files, capturing cross-module relations that static analysis often misses \cite{yang2025kernelgpt}.
For example, opening a file requires not only {\tt fs/open.c}, but also permission checks in {\tt security/}, forming a dispersed dependency that static analysis struggles to identify but remains semantically apparent to an LLM based on the file context.

After these analysis, \toolname{} compiles a comprehensive list of suspicious files that serves as the foundation for subsequent method-level localization. 
This final set includes: (1) files extracted from the crash report, (2) files mapped from syscalls in the test case, and (3) LLM-inferred dependency-linked files.
A concrete illustration is provided in Section~\ref{sec:motivating}, where the final suspicious set includes files from the crash trace (Figure~\ref{fig:motivate}(b)), syscall mappings (Figure~\ref{fig:motivate}(a)), and semantically related dependencies.
To ensure completeness, \toolname{} adopts a conservative strategy that prioritizes expanding the suspicious-file set over early pruning. 
This is crucial, as file-level results directly influence method-level analysis, i.e., any missing files at this stage could compromise the overall localization accuracy. 
By capturing a broader yet semantically grounded file set, \toolname{} lays a robust foundation for fine-grained localization.

\begin{figure}[]
    \centering
    \includegraphics[width=1.0\linewidth]{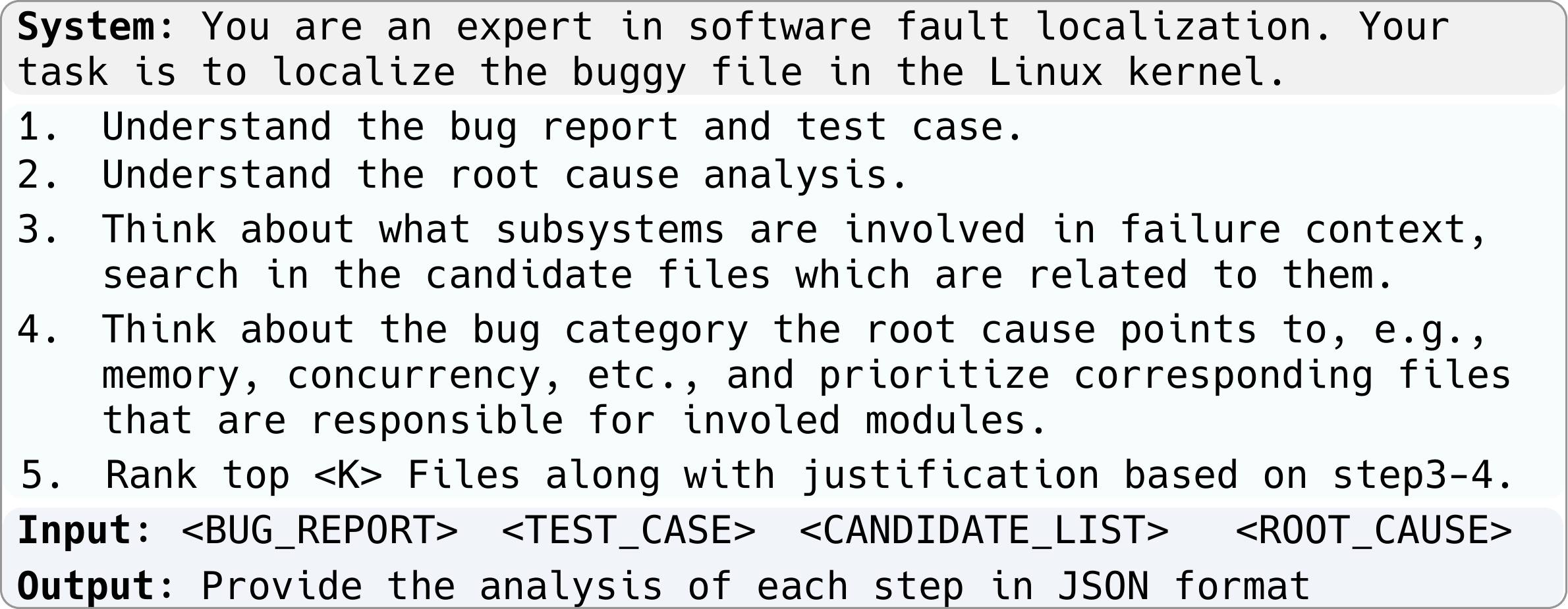}
    
    \caption{Prompt for File-level Fault Localization}
        \label{fig:prompt-localization}
\end{figure}

Compared to scanning the entire kernel codebase, \toolname{}'s file-level localization significantly narrows the search space, improving efficiency and fitting LLM input constraints.
To further prioritize the most relevant files, \toolname{} ranks the suspicious candidates using the contrastive reasoning results described in Section~\ref{subsec:cr}.
Specifically, \toolname{} employs a dedicated file-level localization prompt, illustrated in Figure~\ref{fig:prompt-localization}. 
Following best practices in recent LLM-based techniques, the prompt defines a clear expert role for the model and adopts a CoT prompting strategy to decompose the task into interpretable reasoning steps.
The CoT process begins by instructing the LLM to analyze the bug report, description of inter-file dependency, and the inferred root cause (Steps 1–2). 
Next, the LLM is prompted to summarize the functionality and distinguishing features of each suspicious file. 
Based on this analysis, it ranks the files by estimating their likelihood of being fault-relevant, considering alignment with the root cause and the bug report's contextual clues (Steps 3–5).
The resulting Top-K suspicious files form the input for method-level localization stage, ensuring finer-grained analysis builds on a focused candidate set.

\subsection{Method-Level Fault Localization}
\label{subsec:mlfl}
\input{tables/defect_feature.tex}
While file-level localization significantly reduces the search space, each kernel file still contains about 55 methods on average, making method-level localization computationally demanding.
To further refine the search space, \toolname{} applies method-level filtering by exploiting method-specific characteristics to pinpoint likely faulty methods within suspicious files. 
This filtering consists of two complementary strategies: (1) syntactic-aware filtering and (2) semantics-aware filtering.
Syntactic-aware filtering examines the static code structure to exclude methods exhibiting simple or low-risk patterns, effectively pruning unlikely candidates. 
Semantics-aware filtering then leverages information from bug reports and file-level rankings to evaluate the functional relevance of the remaining methods, enabling more precise candidate selection.
The following sections provide a detailed explanation of each filtering strategy.

\textbf{Syntactic-Aware Filtering.}
\toolname{} employs the widely-used multilayer perceptron (MLP)~\cite{rosenblatt1958perceptron, sun2020learning} to train a lightweight multi-task
defect prediction model to identify and retain highly suspicious methods. 
Using defect prediction models for candidate filtering is a well-established practice in fault localization research~\cite{sohn2017fluccs}. 
Following this practice, we collect historical kernel defect data and extract method-level features to train a model specialized for kernel methods. 
Table~\ref{tab:method_features} summarizes the selected features, including their categories, names, and descriptions. 
These features have been widely adopted and validated in prior defect prediction studies~\cite{kapur2018estimating, pascarella2019fine, nagappan2006mining, kim2015remi}.
Specifically, \toolname{} constructs a labeled dataset comprising historically faulty and non-faulty kernel methods, each represented by a feature vector based on the metrics detailed in Table~\ref{tab:method_features}. 
We then train a multi-task multilayer perceptron model~\cite{rosenblatt1958perceptron, sun2020learning} to distinguish faulty methods from non-faulty ones. 
Applying this model to the methods within the Top-K suspicious files identified by \flfile{}, each method’s feature vector is scored for defect likelihood.  \toolname{} filters methods based on these scores, retaining only those with high likelihood (i.e., the top 80\%) for further semantic analysis.

\textbf{Semantics-Aware Filtering.}
The second filtering stage refines the set of suspicious methods from a functional semantics perspective.
For each candidate method, \toolname{} prompts an LLM to generate a concise summary describing the method’s functionality, based on its name and signature.
These summaries provide semantically meaningful representations that go beyond syntactic structure and are particularly useful for assessing the method’s relevance to the observed bug.
\toolname{} then evaluates each method summary in the context of the bug report and the file-level localization results.
The bug report are typically used as semantic anchors by providing critical keywords, error descriptions, and symptoms, etc. 
The file-level ranking provides a prior suspicion score, reflecting the likelihood that a file contains the faulty method.
To balance precision and efficiency, \toolname{} 
applies a proportion-based candidate selection strategy, guided by file rankings and informed by empirical distributions from prior studies~\cite{zhou2025benchmarking}.
Specifically, \toolname{} allocates more method candidates to higher-ranked files based on their estimated suspiciousness, where 60\% of the candidate methods from the Top-1-3 file, 30\% from files ranked 4–5, and the remaining 10\% to the other files. If prior files lack sufficient methods, the missing candidates will be supplemented from subsequent files.

\begin{figure}[]
    \centering
    \includegraphics[width=1.0\linewidth]{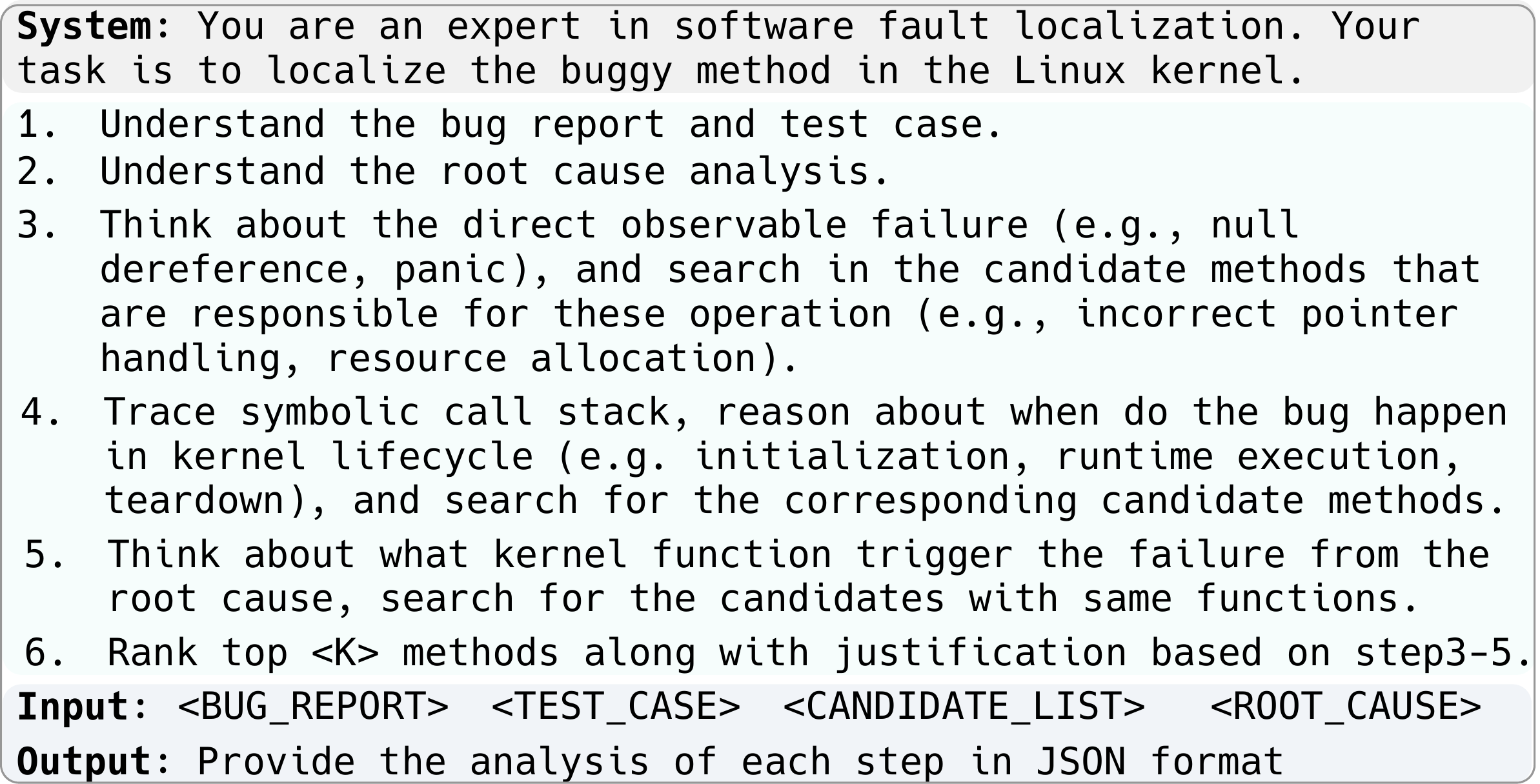}
    
    \caption{Prompt for Method-level Fault Localization}
        \label{fig:prompt-localization-method}
\end{figure}

\toolname{} then ranks the filtered suspicious methods by integrating the contrastive reasoning results described in Section~\ref{subsec:cr}.
To enable effective method-level fault localization, \toolname{} designs a specialized prompt, as illustrated in Figure~\ref{fig:prompt-localization-method}.
While the prompt structure follows the same CoT reasoning framework used at the file level, it is tailored to capture fine-grained method-level semantics and their alignment with the inferred root cause.
Specifically, the prompt guides the LLM through three stages:
The LLM is first instructed to analyze the bug report and contrastive root cause explanation, as well as the LLM-generated summary of each method’s functionality.
For each method, the LLM is asked to evaluate whether its functionality logically aligns with the identified root cause.
This step emphasizes semantic reasoning over structural patterns and requires the LLM to infer whether the method could plausibly implement or affect the faulty behavior.
Based on this alignment, \toolname{} instructs the LLM to assign a suspicion score to each method and output the Top-K most likely faulty methods.

This method-level ranking mechanism enables \toolname{} to connect high-level fault descriptions derived from test behavior to specific code-level implementations.
By explicitly reasoning over the alignment between root causes and method semantics, \toolname{} ensures precise, interpretable, and actionable fault localization.

%% file: tables/defect_feature.tex
\begin{table}[t]
\centering
\caption{Defect Prediction Features of Kernel Method} 
\label{tab:method_features}
\resizebox{\columnwidth}{!}{
\begin{tabular}{cll}
\toprule
\textbf{Category} & \textbf{Feature} & \textbf{Description} \\
\midrule
\multirow{4}{*}{\makecell{Structural\\ Complexity}} 
& Parameters & Parameter numbers of the given method. \\
& Cyclomatic\cite{nagappan2006mining}& Linearly independent paths in the given method. \\
& Halstead Effort\cite{halstead1977elements} & Workload metric based on operator statistics. 
\\
& Max X Depth & Maximum depth of X construct in the method.
\\
\midrule
\multirow{3}{*}{\makecell{Call\\ Relationship}} 
& Call Depth & Maximum local call depth of the given method. 
\\
& In-Degree & Methods that call the given method. \\
& Out-Degree & Method calls in the given method. \\
\midrule
\multirow{3}{*}{\makecell{Code\\ Quality}} 
& Change Time & Time since last modification before versioning.
\\
& Global Variables & Global variables used in the given method.\\
& Static Variables & Static variables used in the given method.\\
\bottomrule
\end{tabular}
}
\end{table}


%% file: 5_evaluation.tex
\section{Evaluation}
\label{sec:evaluation}
\newcommand{\best}[1]{\underline{\textbf{#1}}}

\newcommand{\toolwoCR}{\toolname$_{\textit{w/o}}^{\textit{CR}}$}
\newcommand{\toolwoHCA}{\toolname$_{\textit{w/o}}^{\textit{HCA}}$}

\newcommand{\toolwoSemF}{\toolname$_{\textit{w/o}}^{\textit{SF}}$}
\newcommand{\toolwofi}{\toolname$_{\textit{w/o}}^{\textit{FI}}$}
\newcommand{\toolwomu}{\toolname$_{\textit{w/o}}^{\textit{MU}}$}

To systematically assess the effectiveness of our fault localization technique, we formulate five research questions:

\begin{itemize}[leftmargin=*]
    \item \textbf{RQ1 (File-Level Localization):} How effective is \toolname{} in identifying faulty source files within the Linux kernel?
    
    \item \textbf{RQ2 (Method-Level Localization):} How effective is \toolname{} in pinpointing the exact faulty methods within the Linux kernel?
    

    \item \textbf{RQ3 (Efficiency Analysis):} How efficient is \toolname{} compared with baselines, in terms of runtime and token cost?

    \item \textbf{RQ4 (Ablation Study):} What are the effectiveness and costs of individual components in \toolname{}?

    \item \textbf{RQ5 (Generalizability to non-kernel systems):} How does \toolname{} perform on other (non-kernel) systems ?
    
\end{itemize}



    


\input{5_0_setup}
\subsection{Results and Analysis}

\subsubsection{RQ1: File-Level FL Effectiveness}

\input{tables/fl_file.tex}

The results of RQ1 are shown in Table~\ref{tab:rq1}.
Across \dataall{}, LLM-based techniques (\toolname{}, \agentless{}, and \linuxfl{}) substantially outperform the IR-based technique \irfl{}, achieving improvements of 551\% to 724\% in \textit{Top@1}. 
Among them, \toolname{} consistently delivers the best results.
Compared to \agentless{}, \toolname{} improves \textit{Top@1}--\textit{Top@10} by 4.50\%--12.03\%; against \linuxfl{}, by 18.08\%--26.07\%.
In terms of ranking metrics, \toolname{} achieves a \textit{MRR} of 0.67 and a \textit{MFR} of 2.98, outperforming both \agentless{} (MRR: 0.63, MFR: 3.66) and \linuxfl{} (MRR: 0.55, MFR: 4.34). 
These results demonstrate \toolname{}'s effectiveness, setting a new state-of-the-art.

To evaluate explicit fault evidence’s impact, we test on \datagoodf{} and \databadf{}. 
On \datagoodf{} (with faulty files mentioned), all methods perform better, and \toolname{} performs best by leveraging these direct clues. This aligns with expectations, as LLMs can leverage report mentions or stack traces to directly pinpoint likely fault locations. On \databadf{} (without faulty files), performance drops for all techniques. Yet, \toolname{} still leads, surpassing \linuxfl{} by 50\% in \textit{Top@1} and both baselines by 62.46\%--550.25\% in \textit{Top@3} and 49.98\%--500.16\% in \textit{Top@5}. Notably, \agentless{} fails completely in \textit{Top@1} due to its reliance on surface-level report-code correlations, which fails without explicit mentions. This contrast underscores \toolname{}'s strength in reasoning with indirect evidence.

Statistical analysis confirms \toolname{}'s significance: Friedman test~\cite{demšar2006statistical} shows significant differences ($\chi^2 = 23.512$, $p < 0.001$), with post-hoc Wilcoxon tests revealing \toolname{} outperforms \agentless{} ($W = 584.0$, $p = 0.0002$) and \linuxfl{} ($W = 864.5$, $p < 0.001$). These findings confirm that \toolname{} provides statistically significant improvements in file-level FL effectiveness.

\subsubsection{RQ2: Method-Level FL Effectiveness}

\input{tables/fl_method.tex}

Table~\ref{tab:rq2} shows RQ2 results. \irflmethod{} and \crashlocator{} perform poorly (0\% and 0.85\% \textit{Top@5}), so we omit them from further analysis.


As shown in Table~\ref{tab:rq2}, \toolname{} consistently outperforms LLM-based baselines in method-level fault localization, improving \textit{Top@1}--\textit{Top@10} by 21.04\%--56.85\% and achieving MRR=0.43, MFR=5.61 compared to baselines (MRR: 0.31--0.35, MFR: 6.49--6.94). Though method-level performance is lower than file-level due to finer granularity,
\toolname{} maintains a clear lead, indicating its strong capability in effectively leveraging contextual information and reasoning about the root causes of kernel faults.

On \datagoodm{} (with faulty methods explicit), \toolname{} leads across all metrics, improving \textit{Top@1} by 25.48\%--56.09\%, \textit{Top@3} by 13.25\%--28.75\%, and \textit{Top@5} by 8.91\%--17.85\%. On the more challenging \databadm{}, \toolname{} still achieves a 66.67\%--359.09\% gain in \textit{Top@1} (5.05\% absolute), with notable improvements in \textit{Top@3}--\textit{Top@10} (112.50\%--240.00\%). It ranks the correct method in \textit{Top@10} for 34.34\% of cases. This result demonstrates the robustness and reasoning capability of our contrastive reasoning-enhanced LLM approach in handling incomplete or indirect fault evidence.

Finally, a Friedman test reveals a statistically significant difference in method-level performance across the evaluated approaches ($\chi^2 = 43.24$, $p < 0.001$). Follow-up Wilcoxon signed-rank tests confirm that \toolname{} significantly outperforms \agentless{} ($W = 1077.5$, $p < 0.001$) and \linuxfl{} ($W = 965.0$, $p < 0.001$), indicating our approach's effectiveness in advancing method-level FL for the Linux kernel.

\subsubsection{RQ3: Efficiency Analysis}
\label{rq3}
Columns 1-7 of Table~\ref{tab:time_comparison} show runtime overhead for \toolname{} and baselines.
The Total column reports the overall runtime, and the CR and HCA columns report the time spent by \toolname{} in Contrastive Reasoning (CR) and Hierarchical Context Analysis (HCA), respectively.
Since \flfile{} does not involve method-level localization, the HCA for \flfile{} includes only the file-level localization time.
Based on this table, we can find that \toolname{} is substantially more efficient than the baselines at both \flfile{} and \flmethod{}.
At the more time-consuming \flmethod{}, \toolname{} requires only 332s per case, and achieves speedups of 4.05$\times$ and 1.7$\times$ over \linuxfl{} and \agentless{}, respectively.
Moreover, the \flmethod{} is significantly more time-consuming than \flfile{}, indicating that finer-grained analysis incurs higher overhead.

\input{tables/llmcost.tex}

\begin{figure}[tbp]
  \centering
  \begin{subfigure}[b]{0.45\linewidth}
    \includegraphics[width=\linewidth]{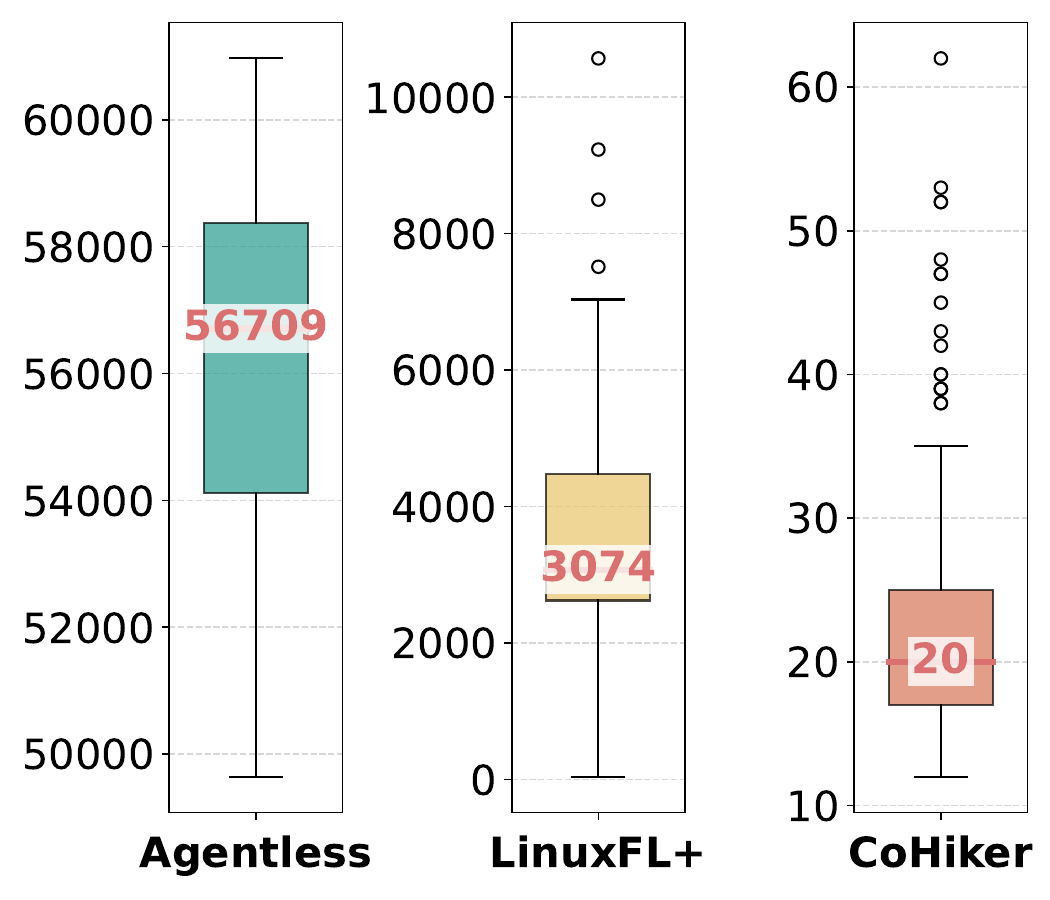}
    \caption{Candidates size at \flfile{}}
    \label{fig:candidates-diff-file}
  \end{subfigure}
  \hspace{0.02\linewidth}  
  \begin{subfigure}[b]{0.45\linewidth}
    \includegraphics[width=\linewidth]{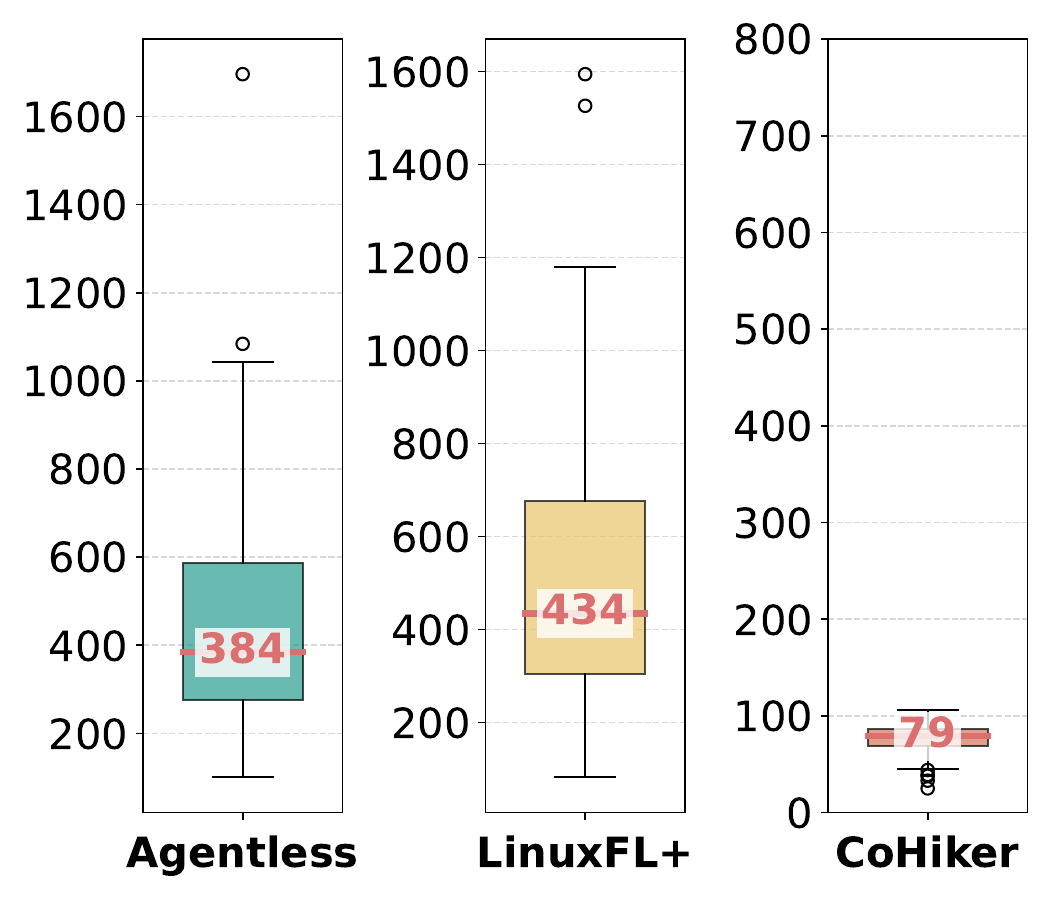}
    \caption{Candidates size at \flmethod{}}
    \label{fig:candidates-diff-method}
  \end{subfigure}
  \caption{Candidates Size of LLM-based Techniques}
  \label{fig:candidates-diff}
\end{figure}

While LLM-based techniques demonstrate promising results compared to traditional approaches, they often incur substantial token usage.
To assess this overhead, we analyzed token consumption of different LLM-based techniques.
As shown in Table~\ref{tab:llm_cost}, \toolname{} achieves the most efficient token usage. 
In comparison, \linuxfl{} consumes 8.84$\times$ more tokens at \flfile{} and 4.46$\times$ more at \flmethod{}. 
\agentless{} incurs the highest overhead, using 28.9$\times$ and 21.92$\times$ more tokens at \flfile{} and \flmethod{}, respectively.
This efficiency stems from \toolname{}'s candidate identification module, which effectively narrows the candidate set before LLM invocation. We further analyzed the size of the candidate set sizes across LLM-based baselines. The results are shown in Figures \ref{fig:candidates-diff-file} and \ref{fig:candidates-diff-method}.

From the two figures, 
it's clear that \toolname{} examines the smallest candidate set at both \flfile{} and \flmethod{} (without sacrificing accuracy).
Especially in \flfile{}, \toolname{} achieves
a median of only 20 candidates, compared to 56,709 (2,835× more) for \agentless{} and 3,074 (153× more) for \linuxfl{}. 
This drastic reduction in input size lowers token usage and enables the LLM to focus on relevant contextual information rather than processing excessive noise.
Specifically, \toolname{} (using \deepseek{}) incurs a cost of approximately \$0.0256 and \$0.0771 per case at the file and method levels, respectively, which is acceptable for practical usage.


\input{tables/runtime}

\subsubsection{RQ4: Ablation Study}
\label{rq3}

RQ4 evaluates the contributions of \toolname{}'s components and LLM reasoning through ablation studies from two perspectives.
First, from the perspective of components, we designed two variants. 
\toolwoCR{} removes CR and infers potential root causes using only the bug report and failing tests. 
\toolwoHCA{} removes HCA(i.e., Syscall Analysis and Dependencies Expansion in \flfile{}, and Syntactic/Semantics Filtering in \flmethod{}), while retaining the two stage localization order(file then method). 
Second, from the perspective of LLM reasoning capability, there are three main stages that rely on LLM.
Thus, we ablate each of these stages and design three variants: \toolwomu{}, \toolwofi{}, and \toolwoSemF{}.
Specifically, \toolwomu{} removes test-case mutation in CR and replaces it with the syntax mutations in Syzkaller~\cite{syzkaller}.
\toolwofi{} removes Syscall Analysis and Dependency Expansion in \flfile{} of HCA, and replaces it with a TF–IDF–based IR model~\cite{salton1988term} to retrieve files lexically related to the syscall, combined with static dependency analysis for file expansion.
\toolwoSemF{} removes Semantic-aware Filtering in \flmethod{} of HCA, and replaces it with the same TF–IDF model to retrieve methods most textually similar to the bug report.
The motivation for designing these variants is to replace LLM-driven semantic reasoning with syntax-based alternatives,
thus assessing the advantage of LLMs.
We then compare the results of these two classes of ablations.

\input{tables/ablation_file.tex}

\noindent \textbf{Contributions of each component:}
Table~\ref{tab:rq3-1} (Rows 2-4, 7-9) presents the results of \toolname{} and its two component ablation variants at \flfile{} and \flmethod{}.
It's evident that removing any component of \toolname{} leads to a decline in localization performance across all metrics.
However, the impact of removing CR is larger than that of removing HCA, indicating that inaccurate root-cause reasoning directly undermines localization accuracy.
To better understand the source of the performance difference, we further analyze which bugs are affected by ablating different components.
Following existing work~\cite{tan2014bug}, we classify all bugs in the dataset into three categories: concurrency bugs (e.g., deadlocks), semantic bugs (e.g., exception handling), and memory bugs (e.g., null-pointer dereference). We refer readers to ~\cite{tan2014bug} for finer-grained definitions.
We then analyze how MRR varies across these categories for each variant, as MRR captures a technique's ability to rank the true faults highly.
The results show that removing CR (i.e., \toolwoCR{}) reduces MRR by 26.51\%, 20.99\%, and 17.97\% for concurrency, semantic, and memory bugs, respectively.
Removing HCA (i.e., \toolwoHCA{}) yields smaller drops of 7.02\%, 20.67\%, and 2.29\% for the same categories.
Based on this, we can find that CR has a substantial impact across all bug types, highlighting the importance of identifying the root cause.
In contrast, HCA has a limited effect on concurrency and memory bugs, which often produce clear crash traces, whereas semantic bugs tend to be more subtle and harder to map to related files.
However, these issues can be mitigated by combining the two components of \toolname{}.

\begin{figure}[t]
    \centering
    \includegraphics[width=1.0\linewidth]{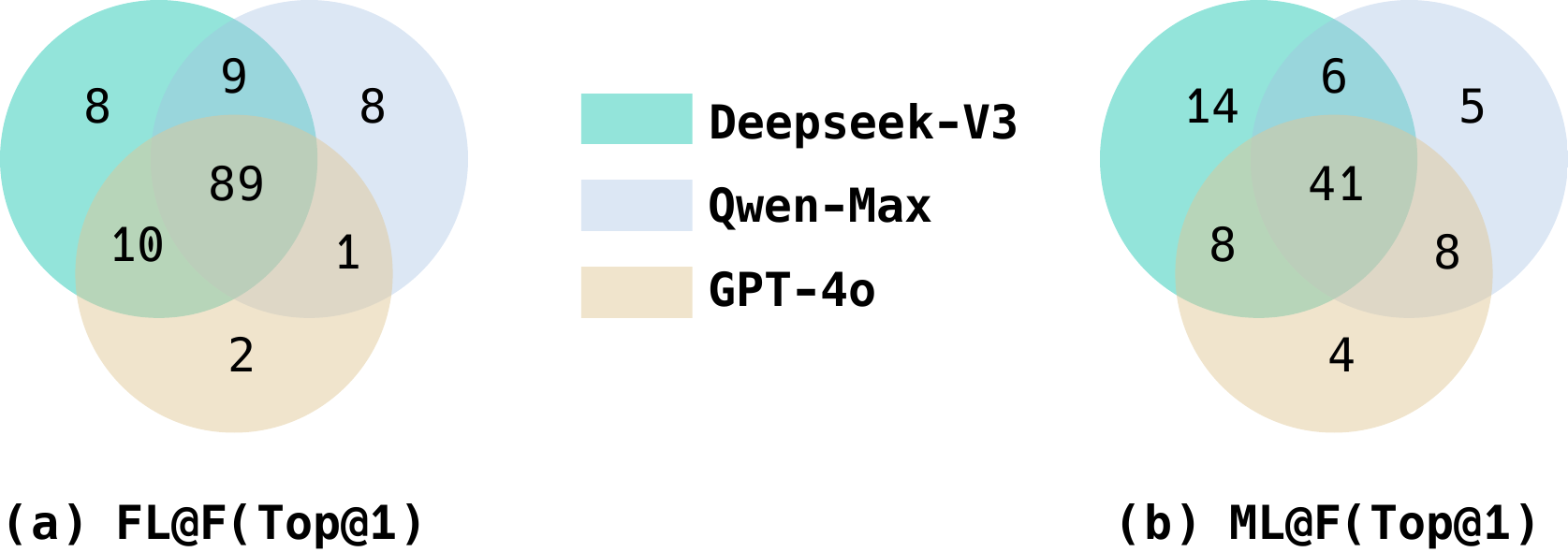}
    \caption{Overlap Analysis with Different LLM Backends}
  \label{fig:llm-diff-top1}
\end{figure}

\noindent \textbf{Contributions of LLM Reasoning:}
Table~\ref{tab:rq3-1} (Rows 5-6, 10-12) presents the result of three \toolname{} variants at \flfile{} and \flmethod{}, where \flfile{} does not include \toolwoSemF{} since it does not involve method-level localization.
The results show that removing any stage that uses LLM leads to a drop in localization performance.
The impact is larger at \flmethod{} than at \flfile{}, indicating that finer-grained localization requires stronger reasoning capability.
Across stages, the overall trends are similar, but removing the Semantic-aware Filtering (\toolwoSemF{}) results in a slightly larger drop, e.g., at \flmethod{}, the Top@3 decreases by 11.71\% for \toolwoSemF{}, compared with drops of 5.72\% and 10.75\% for \toolwomu{} and \toolwofi{}.
This indicates that, at \flmethod{}, semantic similarity matching and filtering based on the LLM-inferred root cause play an important role in localization.
All these findings suggest that although the LLM-based stages can be replaced by traditional techniques (e.g., syntax-based mutation), LLMs still provide additional gains, particularly for finer-grained (e.g., method-level) localization.
Moreover, this motivates a deeper examination of how the LLM backend contributes to \toolname{}'s overall effectiveness.
Thus, besides \deepseek{}, we choose other two widely used LLMs, i.e., \gpt{}~\cite{openai_gpt4o} and \qwenmax{}~\cite{qwen_max}, as the comparative LLM backend. 
Table~\ref{tab:ablation_llm} reports the results of different LLM backends at \flfile{} and \flmethod{}, and Figure~\ref{fig:llm-diff-top1} shows their overlap in correctly locating bugs under Top@1.
Taken together, the results indicate that: 
(1) Different LLM backends induce some performance variation, but the overall effectiveness still surpasses existing techniques, which is largely attributed to \toolname{}'s design, and 
(2) each backend captures different bugs, suggesting that multiple LLM backends can complement each other in practice.

\input{tables/ablation_llm.tex}

Beyond evaluating effectiveness, we also analyzed the efficiency of all five ablated variants to understand accuracy-runtime trade-offs (Columns 8-12 of Table~\ref{tab:time_comparison}).
\toolwomu{} and \toolwofi{} increase runtime, indicating that replacing LLM not only reduces accuracy but also incurs additional overhead, such as generating many invalid mutations.
\toolwoSemF{} similarly brings little runtime benefit. 
In contrast, fully removing CR (\toolwoCR{}) and HCA (\toolwoHCA{}) reduces time cost.
Together with the results in Table~\ref{tab:rq3-1}, we suggest striking a practical balance between these two variants and \toolname{} in practice. 

\subsubsection{RQ5: Generalizability on non-kernel systems}
\label{sec:non_kernel_bench}
Table~\ref{tab:non-kernel-bench} reports the \flfile{} and \flmethod{} results of \toolname{} and the baselines on the dataset of non-kernel systems.
\toolname{} still achieves the best localization performance at both levels, demonstrating the strength and generality of its design.
Another interesting observation is that, compared with the results in Table~\ref{tab:rq3-1}, the localization accuracy on non-kernel systems is much higher at both the file and method levels, indicating that kernel bugs are inherently more difficult to localize.
This highlights the inherent difficulty of kernel fault localization and underscores the significance of our work.

\input{tables/non-kernel-bench}

%% file: 5_0_setup.tex
\subsection{Experimental Setup}

\newcommand{\agentless}{\texttt{Agentless}}
\newcommand{\irfl}{\texttt{Pathidea}}
\newcommand{\irflmethod}{\texttt{FineLocator}}
\newcommand{\linuxflagentless}{$\textit{LinuxFL}+^{*}$}
\newcommand{\crashlocator}{\texttt{Crashlocator}}
\newcommand{\sweagent}{\texttt{SWE-Agent}}
\newcommand{\soapfl}{\texttt{SoapFL}}

\newcommand{\kgym}{\textsc{Kgym} }
\newcommand{\linuxfl}{\texttt{LinuxFL+}}

\subsubsection{Dataset.}
\label{sec:dataset}
To answer RQ1-RQ4, we use a dataset extended from \kgym{}~\cite{mathai2024kgym}, a publicly kernel FL dataset used in previous work~\cite{syzbot}. 
Specifically, it originally consists of 279 Linux kernel bugs with the associated bug reports, test cases, patches, and reproduction instructions. 

Due to platform differences (local Ubuntu 20.04/QEMU 4.2.1 vs. Google Cloud), we reproduced 88 of the original bugs. To ensure representativeness and reliability of the evaluation results, we extended the dataset using \kgym{}'s pipeline, collecting 684 additional bugs from Syzbot~\cite{syzbot} and reproducing 122, yielding 210 total bugs for our final evaluation.
The details of these bugs are presented in Table~\ref{tab:dataset}, where we report the total number of bugs (Total) and the number of bugs that we successfully reproduced (Repro).

\newcommand{\dataall}{$\mathit{D}^{all}$}
\newcommand{\datagoodm}{$\mathit{D}^{\text{MN}}_{M}$}
\newcommand{\databadm}{$\mathit{D}^{\text{UN}}_{M}$}
\newcommand{\datagoodf}{$\mathit{D}^{\text{MN}}_{F}$}
\newcommand{\databadf}{$\mathit{D}^{\text{UN}}_{F}$}

In particular, while manually checking the quality and correctness of the datasets, we found that some bug reports may 
explicitly reference the faulty files or methods, which can significantly reduce the difficulty of fault localization. 
To enable an in-depth analysis of our approach's effectiveness, we further split these bugs into two sub-groups, one of which contains bugs whose bug reports explicitly mentioned the genuine faulty file or faulty method while the other group does not. 
The details are shown in Table~\ref{tab:dataset}. 
To ease the presentation, in the following, we use \datagoodf{} to represent the set of bugs whose bug reports explicitly mention (\textbf{MN}) the faulty files; \databadf{} to represent the set of bugs whose bug reports do not mention (\textbf{UN}) the faulty files; \datagoodm{} to represent the set of bugs whose bug reports explicitly mention the faulty methods; \databadm{} to represent the set of bugs whose bug reports do not mention the faulty methods.

To answer RQ5, we evaluate whether \toolname{} remains effective on non-kernel systems, since its core ideas (contrastive reasoning and hierarchical context analysis) are general and applicable to other software systems. We therefore experiment on SWE-bench-lite \cite{jimenez2024swebench}, a widely used dataset for fault localization on non-kernel systems.
In particular, SWE-bench-lite contains 300 bugs from 11 popular open-source projects (e.g., Django, sympy, matplotlib), each paired with bug reports, bug-triggering test cases, repair patches, and instructions for reproduction, which is similar to our kernel setup, but additionally includes coverage information that is often unavailable for kernel bugs.

\input{tables/dataset.tex}

\input{tables/baselines}

\subsubsection{Baselines.}
\label{sec:baselines}
To comprehensively evaluate the effectiveness of \toolname{}, we compare it with seven fault localization techniques across both kernel and non-kernel scenarios.
Table~\ref{tab:baseline} summarizes all compared techniques, where the Kernel-Specific column indicates whether the technique is designed for kernel fault localization, while the remaining two columns describe the datasets and the localization granularities used in our experiments (determined by the design of each technique).

To answer RQ1-RQ4, we compare \toolname{} against the former five fault localization techniques in Table~\ref{tab:baseline}.
Among these techniques, \linuxfl{}~\cite{zhou2025benchmarking} is the latest 
and the only technique designed for kernel bugs, whereas the others target general-purpose software.
Specifically, 
\crashlocator{}~\cite{wu2014crashlocator} performs localization by analyzing crash stack traces, 
\irfl{}~\cite{chen2021pathidea} and \irflmethod{}~\cite{zhang2019finelocator} are two representative IR-based FL techniques, while \agentless~\cite{xia2024agentless} is the state-of-the-art LLM-based FL technique that represents the latest advance of LLMs in the FL domain~\cite{xia2024agentless}. 
We did not evaluate \sweagent{} and \soapfl{} in RQ1-RQ4, because \sweagent{} serves as the underlying agent framework of \linuxfl{}, which is already evaluated, whereas \soapfl{} requires coverage information that is not always available in kernel.

To answer RQ5, we compare \toolname{} against the latter three localization techniques in Table~\ref{tab:baseline}.
Here, we did not evaluate all techniques, since their effectiveness on non-kernel systems has been extensively studied~\cite{kang2024quantitative, qin2025s}.
Therefore, we select the state-of-the-art \agentless{} and the kernel-specific technique \linuxfl{} as representatives.
However, since \linuxfl{} builds on \sweagent{} by adding kernel-specific information (e.g., mailing-list context) that is unavailable in the non-kernel setting, we use \sweagent{} instead.
Moreover, we additionally compare against \soapfl{}, which identifies suspicious methods through coverage analysis of failing tests. 
We include it here because coverage has never been used for kernel fault localization, making it a meaningful baseline here.

Note that we exclude traditional techniques (e.g., SBFL~\cite{zou2019empirical, wen2019historical}, MBFL~\cite{jia2010analysis, andrews2006using}) because: 
(1) They cannot be effectively applied to kernel bugs (as discussed in Section~\ref{sec:introduction}), and (2) for non-kernel systems, existing LLM-based techniques (e.g., \agentless{}) have been shown to outperform traditional approaches.
For all the compared techniques, we used their default setting and adhered to their original design, as shown in the last two columns of Table~\ref{tab:baseline}.
For techniques not originally designed for kernel bugs, we apply the necessary adaptations, which will be detailed in Section~\ref{sec:implementation}.

\subsubsection{Metrics}

To measure the performance of different techniques in localizing bugs, we adopted three metrics that are commonly used in existing FL studies~\cite{voorhees1999trec, li2019deepfl, zou2019empirical}. The details of them are presented as follows:

\vspace{-1pt}

\begin{itemize}[leftmargin=*]
\item \textit{Top@n}: The ratio of kernel bugs whose faulty file/method appears in the top-n ranked candidates (for $n = 1, 3, 5, 10$).
\item \textit{Mean Reciprocal Rank (MRR)} \cite{voorhees1999trec}: The average reciprocal rank of the ground-truth faulty file/method across all kernel bugs, where higher values indicate better ranking performance.
\item \textit{Mean First Rank (MFR)} \cite{li2021fault}: Given the best rank of the faulty file/method (if involving multiple ones) for a kernel bug, MFR is the average value of the best ranks across all faulty files/methods.

\end{itemize}
\vspace{-2pt}

\subsubsection{Implementation and Configuration}

\label{sec:implementation}

\newcommand{\deepseek}{\texttt{Deepseek-V3}}

\newcommand{\gpt}{\texttt{GPT-4o}}

\newcommand{\qwenmax}{\texttt{Qwen-Max}}

We use Python for implementation, with \deepseek{} \cite{deepseek_v3} as the core LLM backend.
The output limit $K$ is set to 10, aligning with developer preferences \cite{parnin2011automated}. The LLM temperature is 0 to ensure reproducibility.

For syntactic-aware filtering , we follow prior work and train a lightweight defect prediction model , i.e., multilayer perceptron (MLP) model, to filter low-risk methods. 
Specifically, we collect 1,000 additional defective kernel bugs from the Syzbot project (not in our dataset) and train a lightweight multi-task learning defect prediction model using a MLP architecture. It takes a 10-dimensional vector as input, where each dimension represents a normalized value of a structural feature (detailed in Table \ref{tab:method_features}). Training is conducted over 200 epochs with Adam ($lr=1e-3$).
We then filter the bottom 20\% of candidates.
For the semantic-aware filtering, the number of total candidate methods is constrained by the LLM’s context length.
In our setting, a 16k-token context budget allows summarizing approximately 80 methods per bug, using the proportion-based selection strategy from Section~\ref{sec:approach}.

To adapt \toolname{} to SWE-bench-lite, we introduce three modifications.
First, in contrastive reasoning, we mutate function calls in the failing tests (analogous to syscalls in the kernel).
To ensure the validity of mutations, we use static analysis to extract  function signatures and provide them as constraints.
In cases where function calls may be absent, we follow prior work (DIWI~\cite{chen2019compiler}) to perform minimal-expression mutations, such as modifying operators.
Second, in syntactic-aware filtering, we retrain the defect prediction model using 1,000 failing tests from SWE-bench\cite{jimenez2024swebench}.
Finally, in all the LLM-based components, we simply replace kernel-specific context with project-specific context while retaining original design.

We reuse \linuxfl{}'s public implementation with SWEAgent~\cite{yang2024swe}, and faithfully reproduce other baselines by closely following the methodologies described in their respective publications. 
Note that \crashlocator{} was only evaluated on a subset of our dataset, i.e., 123 out of 210 bugs, due to environment incompatibility issues as the Linux kernel bugs in the dataset span hundreds of kernel versions. For \soapfl{}, we directly reuse its publicly released implementation.

We have released all our implementations for replication at \cite{toolrepo}.

\vspace{-2pt}

%% file: tables/dataset.tex
\begin{table}[]
        \caption{Details of the Used Datasets}

\begin{tabular}{@{}c|cccc@{}}
\toprule
Source    & Total & Repro & \textsc{MN} / \textsc{UN} (file) & \textsc{MN} / \textsc{UN} (method) \\ \midrule
Kgym & 279   & 88           & 79/9                             & 50/38                              \\
Syzbot        & 684   & 122          & 83/40                            & 61/61                              \\ \midrule
Summary       & 963   & 210          & 161/49                           & 111/99                             \\ \bottomrule
\end{tabular}
\label{tab:dataset}
\end{table}

%% file: tables/baselines.tex
\begin{table}[]
    \caption{Details of the Compared Baselines}
    \label{tab:baseline}
    \resizebox{\columnwidth}{!}{
\begin{tabular}{@{}c|c|cc|cc@{}}
\toprule
\multirow{2}{*}{Techniques} & \multirow{2}{*}{\makecell{Kernel\\Specific}} & \multirow{2}{*}{\makecell{Kernel Dataset\\ (RQ1-RQ4)}} & \multirow{2}{*}{\makecell{Non-kernel Dataset\\(RQ5)}} & \multicolumn{2}{c}{FL level} \\ \cmidrule(l){5-6} 
                            &                                  &                                                        &                                                       & \flfile{}    & \flmethod{}   \\ \midrule
\irfl{}                     & \ding{55}                        & \ding{51}                                              & \ding{55}                                             & \ding{51}    & \ding{55}     \\
\irflmethod{}               & \ding{55}                        & \ding{51}                                              & \ding{55}                                             & \ding{55}    & \ding{51}     \\
\crashlocator{}             & \ding{55}                        & \ding{51}                                              & \ding{55}                                             & \ding{55}    & \ding{51}     \\
\linuxfl{}                  & \ding{51}                        & \ding{51}                                              & \ding{55}                                             & \ding{51}    & \ding{51}     \\
\agentless{}                & \ding{55}                        & \ding{51}                                              & \ding{51}                                             & \ding{51}    & \ding{51}     \\
\sweagent{}                 & \ding{55}                        & \ding{55}                                              & \ding{51}                                             & \ding{51}    & \ding{51}     \\
\soapfl{}                   & \ding{55}                        & \ding{55}                                              & \ding{51}                                             & \ding{51}    & \ding{51}     \\ \bottomrule
\end{tabular}}
\end{table}

%% file: tables/fl_file.tex
\begin{table}[]
      \footnotesize
    \caption{File-level Fault Localization Result (RQ1)}
\resizebox{\columnwidth}{!}{
\begin{tabular}{@{}c|c|cccc|cc@{}}
\toprule
Data                        & Technique   & \textit{Top@1}   & \textit{Top@3}   & \textit{Top@5}   & \textit{Top@10}  & \textit{MRR} & \textit{MFR} \\ \midrule
\multirow{4}{*}{\dataall}   & \irfl      & 6.7\%          & 13.4\%         & 15.78\%        & 34.92\%        & 0.11         & 8.06         \\
                            & \agentless & 52.86\%        & 70\%           & 75.24\%        & 81.43\%        & 0.63         & 3.66         \\
                            & \linuxfl   & 43.6\%         & 63.81\%        & 70.95\%        & 72.86\%        & 0.55         & 4.34         \\
                            & \toolname  & \best{55.24\%} & \best{77.62\%} & \best{84.29\%} & \best{89.05\%} & \best{0.67}  & \best{2.98}  \\ \midrule
\multirow{4}{*}{\datagoodf} & \irfl      & 8.75\%         & 15.63\%        & 18.75\%        & 42.5\%         & 0.14         & 7.93         \\
                            & \agentless & 68.52\%        & 89.51\%        & 95.68\%        & 98.15\%        & 0.80         & 1.79         \\
                            & \linuxfl   & 55.56\%        & 77.78\%        & 84.57\%        & 85.8\%         & 0.67         & 2.98         \\
                            & \toolname  & \best{69.75\%} & \best{92.59\%} & \best{98.15\%} & \best{98.77\%} & \best{0.81}  & \best{1.63}  \\ \midrule
\multirow{4}{*}{\databadf}  & \irfl      & 0              & 6.12\%         & 6.12\%         & 10.2\%         & 0.03         & 10.41        \\
                            & \agentless & 0              & 4.08\%         & 6.12\%         & 26.53\%        & 0.05         & 9.84         \\
                            & \linuxfl   & 4.08\%         & 16.33\%        & 24.49\%        & 30.61\%        & 0.12         & 8.80         \\
                            & \toolname  & \best{6.12\%}  & \best{26.53\%} & \best{36.73\%} & \best{55.1\%}  & \best{0.20}  & \best{7.39}  \\ \bottomrule
\end{tabular} }
\label{tab:rq1}
\end{table}

%% file: tables/fl_method.tex
\begin{table}[t]
    \footnotesize
    \caption{Method-level Fault Localization Result (RQ2)}
    \resizebox{\columnwidth}{!}{

\begin{tabular}{@{}c|c|cccc|cc@{}}
\toprule
Data                          & Technique            & \textit{Top@1}   & \textit{Top@3}   & \textit{Top@5}   & \textit{Top@10}  & \textit{MRR} & \textit{MFR} \\ \midrule
\multirow{3}{*}{\dataall{}}   & \agentless{}        & 20.95\%        & 37.14\%        & 43.81\%        & 54.29\%        & 0.31         & 6.94         \\
                              & \linuxfl{}          & 24.76\%        & 43.33\%        & 48.1\%         & 53.81\%        & 0.35         & 6.49         \\
                              & \toolname{}         & \best{32.86\%} & \best{52.86\%} & \best{58.57\%} & \best{65.71\%} & \best{0.43}  & \best{5.61}  \\ \midrule
\multirow{3}{*}{\datagoodm{}} & \agentless{}        & 36.94\%        & 65.77\%        & 75.68\%        & 93.69\%        & 0.54         & 3.94         \\
                              & \linuxfl{}          & 45.95\%        & 74.77\%        & 81.89\%        & 89.19\%        & 0.62         & 3.36         \\
                              & \toolname{}         & \best{57.66\%} & \best{84.68\%} & \best{89.19\%} & \best{93.69\%} & \best{0.71}  & \best{2.74}  \\ \midrule
\multirow{3}{*}{\databadm{}}  & \agentless{}        & 3.03\%         & 5.05\%         & 8.08\%         & 10.1\%         & 0.05         & 10.35        \\
                              & \linuxfl{}          & 1.1\%          & 8.08\%         & 10.1\%         & 14.14\%        & 0.05         & 10.03        \\
                              & \toolname{}         & \best{5.05\%}  & \best{17.17\%} & \best{24.24\%} & \best{34.34\%} & \best{0.12}  & \best{8.87}  \\ \bottomrule

\end{tabular}
}
\label{tab:rq2}
\end{table}


%% file: tables/llmcost.tex
\begin{table}[tbp]
    \small
    \caption{Token Cost for LLM-based Techniques (RQ3)} 
    \label{tab:llm_cost}
    \centering
    \resizebox{\linewidth}{!}{  
    \begin{tabular}{c|ccc}
        \toprule
        \multirow{2}{*}{FL level} & \multicolumn{3}{c}{Token cost per case. (Input / Output / Total)} \\ \cmidrule(l){2-4} 
                                  & \linuxfl    & \agentless               & \toolname                \\ \midrule
        \flfile                   & 94.3K / 3.61K / 97.9K         & 0.32M / 2.72K / 0.32M    & 10.5K / 0.57K / 11.07K   \\
        \flmethod                 & 0.11M / 4.01K /0.11M          & 0.54M / 4.4K / 0.54M     & 20K / 4.64K / 24.63K     \\ 
        \bottomrule
    \end{tabular}
    }
\end{table}

%% file: tables/runtime.tex
\begin{table*}[]
\centering
\caption{Runtime Cost Comparision between Different Techniques (RQ3\&4)}
\label{tab:time_comparison}
\resizebox{\linewidth}{!}{
\begin{tabular}{@{}c|ccc|ccc|ccccc@{}}
\toprule
\multirow{2}{*}{FL Level} & \multicolumn{3}{c|}{\toolname{}} & \multicolumn{3}{c|}{Baselines}      & \multicolumn{5}{c}{Ablation mutants of \toolname{}}                   \\ \cmidrule(l){2-12} 
                          & CR        & HCA     & Total    & \irfl{} & \linuxfl{} & \agentless{}  &  \toolwoCR{} & \toolwoHCA{}  & \toolwomu{} & \toolwofi{} & \toolwoSemF{} \\ \midrule
\flfile{}                 & 138.5s     & 10.4s      & \textbf{148.9s}    & 195.8s & 1312.0s                     & 549.6s                                           & 10.4s (- 93.0\%)                      & 144.7s (- 2.8\%)                        & 234.4s  (+ 57.4\%)                     & 149.4s    (+ 0.3\%)                   & 148.9s    (- 0.0\%)                     \\
\flmethod{}               & 138.5s     & 193.5s     & \textbf{332s}      & -     & 1345.5s                     & 566.6s                                            & 193.5s  (- 41.7\%)                       & 160.67s     (-51.6\%)                    & 417.5s    (+ 25.8\%)                      & 332.5s   (+0.2\%)                       & 331.2s    (-0.2\%)                        \\ \bottomrule
\end{tabular}}
\end{table*}

%% file: tables/ablation_file.tex


\begin{table}[t]
\caption{Contribution of each Components (RQ4)} 
    \centering
    \resizebox{\columnwidth}{!}{
\begin{tabular}{@{}c|c|cccc|cc@{}}
\toprule
FL Level                     & Technique    & \textit{Top@1} & \textit{Top@3} & \textit{Top@5} & \textit{Top@10} & \textit{MRR} & \textit{MFR} \\ \midrule
\multirow{5}{*}{\flfile{}}   & \toolname{} & \best{55.24\%}      & \best{77.62\%}       & \best{84.29\%}      & \best{89.05\%}       & \best{0.67}        & \best{2.98}         \\
                             & \toolwoCR{} & 54.29\%      & 73.33\%      & 77.76\%      & 81.9\%        & 0.64         & 3.51         \\
                             & \toolwoHCA{} & 47.14\%      & 69.05\%      & 78.57\%      & 86.67\%       & 0.60         & 3.49         \\ 
                             \cmidrule(l){2-8} 
                             & \toolwomu{} & 53.33\%       & 73.33\%       & 81.43\%      & 87.14\%       & 0.65         & 3.31         \\
                             & \toolwofi{} & 53.81\%      & 73.33\%      & 76.67\%      & 78.10\%       & 0.64         & 3.66         \\ \midrule
\multirow{6}{*}{\flmethod{}} & \toolname{} & \best{32.86\%}      & \best{52.86\%}       & \best{58.57\%}      & \best{65.71\%}       & \best{0.43}         & \best{5.61}         \\
                             & \toolwoCR{} & 26.79\%      & 44.98\%      & 50.24\%      & 58.37\%       & 0.37         & 6.63         \\
                             & \toolwoHCA{} & 27.54\%      & 47.55\%      & 53.43\%      & 65.20\%       & 0.39         & 6.05         \\  \cmidrule(l){2-8} 
                             & \toolwomu{} & 28.09\%      & 47.14\%      & 53.81\%      & 62.38\%       & 0.41         & 5.99         \\
                             & \toolwofi{} & 29.07\%      & 42.11\%      & 49.28\%      & 58.85\%       & 0.38         & 6.11         \\
                             & \toolwoSemF{} & 25.48\%      & 41.15\%      & 46.89\%      & 54.67\%       & 0.34         & 6.68         \\ \bottomrule
\end{tabular}
}
\label{tab:rq3-1}
\end{table}



%% file: tables/ablation_llm.tex


\begin{table}[t]
    \footnotesize
    \caption{Ablation Study on LLM Backends (RQ4)}

\begin{tabular}{c|c|cccc|c}
\toprule
FL level                                        & 
LLMs                             & \textit{Top@1} & \textit{Top@3} & \textit{Top@5} & \textit{Top@10} & \multicolumn{1}{l}{\textit{MRR}} \\ \midrule
\multicolumn{1}{c|}{\multirow{3}{*}{\flfile}}   & 
\multicolumn{1}{c|}{\deepseek{}} & 
\best{55.24\%}    & \best{77.62\%}    & \best{84.29\%}    & \best{89.05\%}     & \best{0.67}                   \\
\multicolumn{1}{c|}{}                           & \multicolumn{1}{c|}{\qwenmax{}}    & 50.95\%           & 71.73\%           & 77.14\%           & 84.29\%            & 0.60                           \\
 \multicolumn{1}{c|}{}                           & \multicolumn{1}{c|}{\gpt{}}      &
 48.57\%           & 72.38\%           & 80.00\%           & 83.33\%            & 0.62                           \\ \midrule
 \multicolumn{1}{c|}{\multirow{3}{*}{\flmethod}} & 
 \multicolumn{1}{c|}{\deepseek{}} & \best{32.86\%}    & \best{52.86\%}    & \best{58.57\%}    & \best{65.71\%}     & \best{0.43}                     \\
\multicolumn{1}{c|}{}                           & 
\multicolumn{1}{c|}{\qwenmax{}}    & 28.57\%           & 42.38\%           & 52.38\%           & 59.05\%            & 0.38                           \\
\multicolumn{1}{c|}{}                           & \multicolumn{1}{c|}{\gpt{}}      &  29.05\%           & 39.52\%           & 50\%              & 58.1\%             & 0.37                          \\ \bottomrule
\end{tabular}
\label{tab:ablation_llm}
\end{table}

%% file: tables/non-kernel-bench.tex
\begin{table}[]
\caption{Fault Localization Result on Non-kernel Systems}
    \resizebox{\columnwidth}{!}{
\begin{tabular}{@{}c|ccccc|cc@{}}
\toprule
FL-Level                & Technique    & \textit{Top@1} & \textit{Top@3} & \textit{Top@5} & \textit{Top@10} & \textit{MRR} & \textit{MFR} \\ \midrule
\multirow{4}{*}{\flfile{}} & \agentless{} & 62.04\%        & 78.83\%        & 83.21\%           & 86.86\%         & 0.715        & 2.91         \\ 
                        & \sweagent{}  & 56.72\%        & 69.40\%        & 73.13\%        & 76.86\%         & 0.637        & 3.32         \\
                        & \soapfl{}    & 63.49\%        & 80.15\%        & 81.34\%        & 81.34\%         & 0.716        & 3.11         \\ 
                        & \toolname{}  & \best{73.36\%} & \best{89.41\%} & \best{94.16\%} & \best{95.62\%}  & \best{0.821} & \best{1.89}  \\ \midrule
\multirow{4}{*}{\flmethod{}} & \agentless{} & 51.62\%        & 58.544\%        & 61.38\%        & 62.20\%         & 0.557        & 5.01    \\
                        & \sweagent{}  & 45.52\%        & 64.18\%        & 68.66\%        & 73.13\%         & 0.554        & 4.4          \\
                        & \soapfl{}    & 49.80\%        & 61.13\%        & 65.18\%        & 67.61\%         & 0.585        & 4.46         \\ 
                        & \toolname{}  & \best{54.37\%} & \best{71.89\%} & \best{77.37\%} & \best{84.30\%}  & \best{0.643} & \best{3.41}  \\ \bottomrule
\end{tabular}
}
\label{tab:non-kernel-bench}
\end{table}

%% file: 7_discussion.tex
\section{Discussion}
\label{sec:discussion}


Like other LLM-based localization methods, \toolname{} may also be affected by data leakage if the studied kernel bugs were used during model pre-training. 
To mitigate this, we evaluated 21 bug reports dated after March 24, 2025, the release date of the \deepseek{} model used in \toolname{}.
The results show that \toolname{} performs similarly on these unseen bugs, indicating that data leakage has minimal impact.
In the future, we will explore additional ways to reduce data leakage and further deepen kernel FL, such as examining performance differences among baselines and improving localization granularity (e.g., toward line-level).


%% file: 6_relatedwork.tex
\section{Related Work}
\label{sec:related_work}

\subsection{Non-LLM-based Fault Localization}
Typical FL techniques include spectrum-based (SBFL), mutation-based (MBFL), information retrieval-based (IRFL), and hybrid techniques.
SBFL is one of the most widely studied FL techniques \cite{zou2019empirical, wen2019historical}, which are broadly applicable to software debugging tasks~\cite{abreu2007accuracy, wong2013dstar}. SBFL calculates suspiciousness scores for each program element according to the test coverage discrepancy between failing and passing runs via a predefined formula, e.g., Ochiai \cite{abreu2007accuracy}, Dstar\cite{wong2013dstar} and Tarantula\cite{jones2005empirical}. Similar to SBFL, MBFL leverages the concept of mutation testing \cite{jia2010analysis} to generates program variants and scores them based on coverage differences~\cite{moon2014ask, papadakis2015metallaxis}.
However, both SBFL and MBFL require runtime coverage data, which is often unavailable in the context of Linux kernel failures. The reason is that execution traces in the kernel are held in the volatile memory, which will be lost upon a system crash \cite{bissyande2012diagnosys}.
IRFL ranks suspicious locations by textual similarity between bug reports and code~\cite{xia2023information, zhang2019finelocator, chen2021pathidea, razzaq2021boostnsift}.
For example, Finelocator \cite{zhang2019finelocator} improves method-level localization by augmenting short method representations with neighbor features for similarity computation.
BoostNSift \cite{razzaq2021boostnsift} improves fault localization performance by enhancing the quality of bug report titles for more accurate relevance measurement.
The most recent Pathidea \cite{chen2021pathidea} incorporates the static call graph to aid the construction of the failing execution path, which further improves the representation of textual features from bug reports. 
Although IRFL techniques can effectively leverage textual information, they often struggle with the complexity of kernel code and the sparsity of symbolic information (e.g., function names) in kernel stack traces. This makes it difficult to accurately match bug reports with relevant code entities, especially in large codebases like the Linux kernel.

\subsection{LLM-based Fault Localization}

LLMs' success has boosted their adoption in fault localization.
Wu et al.~\cite{wu2023large} prompt LLMs with code and error
logs, and instruct the model to identify the buggy lines, while LLMAO~\cite{yang2024large} trains lightweight bidirectional adapters on top of LLMs to produce a suspiciousness score for each line. FuseFL~\cite{widyasari2024demystifying} further feeds the spectrum information and test execution results to LLMs for identifying suspicious code. These approaches, however, are designed for small code snippets but not suitable for large codebases like the Linux kernel due to their reliance on inputting the source code. 

To address LLM input length limitations, AutoFL~\cite{kang2024quantitative} uses function calling capability to inspect covered methods and pinpoint buggy methods.
Similarly, FlexFL \cite{xu2025flexfl} first collects SBFL/IRFL candidates, and then utilizes LLM agents to refine the results by analyzing code structures and bug reports. SoapFL \cite{qin2025s} leverages LLM agents to iteratively narrow down the search space by analyzing test cases' dynamically execution paths.
As a consequence, they are also inapplicable to localize Linux kernel bugs due to the shared limitations, e.g., unavailable code coverage and sparse symbolic information.
The most recent Agentless \cite{xia2024agentless} addresses the above limitations through
a recursive fault localization strategy, which identifies suspicious files by iteratively invoking the LLM for search space reduction, making it extremely costly (\textit{ref.} Table~\ref{tab:llm_cost}).
LinuxFL+\cite{zhou2025benchmarking}, latest research designed for kernel faults, uses LLM agents to identify suspicious files and re-ranks them through incorporating contextual features of code.

Compared to these existing approaches, \toolname{} is specially designed to address the challenges in Linux kernel fault localization. Specifically, it incorporates a contrastive reasoning component for fine-grained root cause identification and search space reduction. It has been proved to be effective in our comprehensive evaluation.

%% file: 8_conclusion.tex
\section{Conclusion}
\label{sec:conclusion}

\noindent
In this work, we present \toolname{}, a novel LLM-based fault localization technique tailored for the Linux kernel.
By introducing a contrastive reasoning component and a hierarchical feature integration framework, \toolname{} addresses key limitations of prior approaches, i.e., \textit{lack of precise root cause analysis} and \textit{ineffective context utilization}.
Our design enables more accurate and efficient fault localization by aligning LLM reasoning with kernel-specific characteristics at both file and method levels.
Through a comprehensive evaluation on 210 real-world kernel bugs, \toolname{} outperforms both traditional and state-of-the-art LLM-based baselines in localization accuracy, while significantly reducing token consumption. Specifically, compared to LLM-based baselines \toolname{} improves \textit{Top@1} by up to 26.70\% at the file level and 56.85\% at the method level, and reduces LLM token usage by up to 8.84$\times$ and 28.9$\times$, respectively. Furthermore, \toolname{} generalizes well to the non-kernel dataset, confirming its applicability across diverse systems.

%% file: 10_acknowledge.tex
\begin{acks}

This work is supported by National Key Research and Development Program of China (No. 2024YFB4506300), and National Natural Science Foundation of China (Nos. 62322208, 12411530122,
6257234, 62202324 and
62472310).
\end{acks}